\documentclass[preprint]{emulateapj}
\usepackage{amsmath}

\def\Q{{\cal Q}}
\def\R{{\cal R}}

\newcommand{\simgt}{\,\hbox{\lower0.6ex\hbox{$\sim$}\llap{\raise0.6ex\hbox{$>$}}}\,}
\newcommand{\simlt}{\,\hbox{\lower0.6ex\hbox{$\sim$}\llap{\raise0.6ex\hbox{$<$}}}\,}
\newcommand{\code}[1]{\texttt{#1}}

\begin{document}

\title{Testing Secondary Models for the Origin of Radio Mini-Halos in Galaxy Clusters}

\author{J.A. ZuHone\altaffilmark{1}, G. Brunetti\altaffilmark{2},
  S. Giacintucci\altaffilmark{3}, M. Markevitch\altaffilmark{1}}

\altaffiltext{1}{Astrophysics Science Division, Laboratory for High Energy Astrophysics, Code 662, NASA/Goddard Space Flight Center, Greenbelt, MD 20771, USA}
\altaffiltext{2}{INAF – Istituto di Radioastronomia, via Gobetti 101, 40129 Bologna, Italy}
\altaffiltext{3}{Department of Astronomy, University of Maryland, College Park, MD, 20742-2421, USA}
 
\begin{abstract}
We present an MHD simulation of the emergence of a radio minihalo in a galaxy
cluster core in a ``secondary'' model, where the source of the
synchrotron-emitting electrons is hadronic interactions between cosmic-ray
protons with the thermal intracluster gas, an alternative to the ``reacceleration model'' where the cosmic ray electrons are reaccelerated by turbulence induced by core sloshing, which we discussed in an earlier work. We follow the evolution of cosmic-ray electron spectra and their radio emission using passive tracer particles, taking into account the time-dependent injection of electrons from hadronic interactions and their energy losses. We find that secondary electrons in a sloshing cluster core can generate diffuse synchrotron emission with luminosity and extent similar to observed radio minihalos. However, we also find important differences with our previous work. We find that the drop in radio emission at cold fronts is less prominent than that in our reacceleration-based simulations, indicating that in this flavor of the secondary model the emission is more spatially extended than in some observed minihalos. We also explore the effect of rapid changes in the magnetic field on the radio spectrum. While the resulting spectra in some regions are steeper than expected from stationary conditions, the change is marginal, with differences in the synchrotron spectral index of $\Delta\alpha\simlt 0.15-0.25$, depending on the frequency band. This is a much narrower range than claimed in the best-observed minihalos and produced in the reacceleration model. Our results provide important suggestions to constrain these models with future observations.
\end{abstract}

\section{Introduction\label{sec:intro}}

A number of relaxed, cool-core clusters are hosts to faint, diffuse
synchrotron radio sources called ``radio minihalos'', with a radius comparable to the size of the cooling region ($r \simlt$~100-300~kpc) and a steep spectrum ($\alpha \simgt 1$;
$S_{\nu} \propto \nu^{-\alpha}$). They arise from cosmic-ray electrons
(hereafter CRe) emitting in the $\sim\mu$G magnetic fields in the
cluster. These are relatively rare sources, with currently only around 10 clusters with confirmed detections. Examples include Perseus \citep{bur92}, A2029
\citep{gov09}, Ophiuchus \citep{gov09,mur10}, RXC J1504.1-0248
\citep{gia11}, and RXJ 1347-1145 \citep{git07}, along with the newest
detections in \citet{gia14a}. The origin of radio mini-halos in cool
core clusters, as well as their possible connection with giant radio
halos, is still unclear \citep[see, e.g.][for a review]{bru14}. Though the clusters
which host these sources typically also have active galactic nuclei,
due to the fact that the radiative timescale of the electrons at the
required energies for the observed emission ($\sim{10^8}$~years) is much
shorter than the time required for these electrons to diffuse across
the cooling region, they alone cannot account for the origin of minihalos.

Most cool-core clusters also possess spiral-shaped cold fronts \citep{ghi10}, which
are believed to be the product of the central cold gas ``sloshing'' in
the cluster potential minimum due to interactions with
subclusters. A number of simulation works have investigated the
importance of this phenomenon \citep[e.g.][]{AM06,zuh10,rod11a}. Most relevant for this work, in \citet[][hereafter ZML11]{zuh11} we showed that such magnetic
field amplification occurs near the cold front surfaces due to the local velocity shear, and that the magnetic field within the core region is stronger than it was previous to the onset of sloshing. 

The relevance of gas sloshing to radio minihalos was first pointed out by \citet{maz08}, who discovered a correlation between sloshing
cold fronts and minihalo emission in two galaxy clusters. A number of 
other examples have since been observed \citep[e.g.][]{hla13,gia14a}. In these cases, the minihalo emission is well-confined to the area on the sky delineated by the cold front surfaces, suggesting a causal connection between sloshing motions and radio minihalos. 

One possible origin for the CRe which produce the observed synchrotron
radiation of minihalos is reacceleration of a lower energy population
of CRe due to turbulence \citep{git02}. In a previous work \citep[][hereafter Z13]{zuh13}, 
we investigated the possibility that radio minihalos are due to turbulent reacceleration in sloshing cluster cores. The sloshing motions produced turbulent velocities of $\delta{v} \sim 100-200~{\rm km~s^{-1}}$ on scales of 10s of kpc within the sloshing region. We showed that the radio emission produced in these simulations was consistent with the properties of observed minihalos where sloshing cold fronts are also found. In particular, the minihalo emission was confined to the core region and possessed a steep spectrum.

There is an alternative perspective on the origin diffuse radio emission in
clusters of galaxies, known as the ``hadronic'' or ``secondary'' model
\citep[see, e.g.][]{pfr04, kes10a, fuj12}. Cosmic-ray protons (hereafter CRp) are believed
to fill the cluster volume, having been accelerated from the thermal
population to relativistic energies by supernovae, AGN, and shocks associated with cosmological structure formation \citep[for a review see][]{bru14} These protons will undergo interactions with the thermal proton population, producing pions which
will decay into secondary products, including cosmic-ray electrons and positions:
\begin{eqnarray}
p_{\rm th} + p_{\rm CR} & \rightarrow & \pi^0 + \pi^+ + \pi^- \\
\pi^\pm & \rightarrow & \mu^\pm + \nu_\mu/\bar{\nu}_\mu \rightarrow
e^\pm + \nu_e/\bar{\nu}_e + \nu_\mu + \bar{\nu}_\mu \nonumber \\
\pi^0 & \rightarrow & 2\gamma \nonumber
\end{eqnarray}
Since the CRp have very long radiative loss times
compared with the age of the cluster, these interactions should
continuously provide a fresh population of CRe at a range of
energies. In this model, it is these CRe that produce the observed
radio emission. These hadronic interactions should also produce a flux
of $\gamma$-rays, but so far no confirmed detections of such emission
from galaxy clusters have been made. Upper limits on the $\gamma$-ray
flux from several experiments \citep[see, e.g.][]{ale12,fer13,pro13} indicate the ratio of the CRp energy to the thermal energy of the gas is at most $\sim$1-2\%. Still,
contrary to the case of nearby giant radio halos, these limits do not
put significant tension on a secondary origin of mini-halos \citep[][]{bru14}. For
example, \citet{zan14} proposed that mini-halos are primarily of
hadronic origin, while giant radio halos experience a transition from
a central hadronic emission to a leptonic emission component in the
external regions due to CRe reacceleration.

In the hadronic model, under relatively quiescent conditions, the electron spectrum will reach a steady-state condition where the energy losses and the gains due to injection balance each other out \citep{sar99}. If the spectrum of CRp is a power law with spectral index $\alpha_p$, at high energies, where the radiative (synchrotron and inverse-Compton) losses dominate, the CRe spectrum has the form $N(E) \propto E^{-\alpha_e}$, which results in a synchrotron emissivity that depends on the properties of the CRp and the plasma as
\begin{equation}\label{eqn:ss_j}
j_\nu \propto n_{\rm th}\epsilon_{\rm CRp}\frac{B^{1+\alpha}}{B^2+B_{\rm CMB}^2}\nu^{-\alpha}
\end{equation}
where $n_{\rm th}$ is the number density of thermal particles, $B$ is the
magnetic field strength, $B_{\rm CMB} \approx 3.25(1+z)^2~\mu$G is the
equivalent magnetic field strength of the CMB, $\alpha$ is the
synchrotron spectral index, and $\epsilon_{\rm CRp}$ is the energy
density of CRp. Under the steady-state assumption and where radiative
losses dominate, $\alpha = \alpha_p/2 \sim 1$, using a canonical value $\alpha_p \sim 2-2.5.$ This implies that for $B \gg B_{\rm CMB}$ the radio emission is roughly independent of the magnetic field strength and for $B \ll B_{\rm CMB}$ it is suppressed. 

Though this spectral index value is consistent with the range
of spectral indices found in minihalo sources, this simplified picture
implies that the synchrotron spectrum should have the same slope everywhere throughout
the minihalo. So far, the few available spatially-resolved
observations of the minihalo spectral slopes indicate that
this is likely not the case \citep{sij93,mur10,git13,gia14b}, although
more detailed analyses are needed. Under stationary conditions, and if diffusion and/or transport of CRp and CRe is not important, this power-law slope should extend to high frequencies, without any steepening (distinct from the turbulent acceleration model, which predicts steepening at high frequencies due to the balance between reacceleration and losses on the CRe).

However, the conditions of cluster cores with minihalos probably do not lend themselves to a such a steady-state assumption, given the existence of cold fronts which betray the presence of sloshing motions. As previously noted (ZML11), sloshing motions can significantly amplify the magnetic field strength on short timescales. \citet{kes10b} demonstrated that a rapidly increasing magnetic field strength will result in a steeper spectrum than the steady-state case (so long as $B \gg B_{\rm CMB}$). In the present case, magnetic field amplification near cold front surfaces would cause spectral steepening of the radio emission moving outward along the spiral arm of the sloshing cold fronts, away from the cluster center.

The same magnetic field amplification has possible implications for the extent of the radio emission as well. \citet{kes10a} predicted that as a result of the strengthening of the magnetic field, the radio emission would have a steep gradient at the front surface, since inside the front $B \gg B_{\rm CMB}$, and outside, $B \ll B_{\rm CMB}$.

The appeal of this model for the steepening of the radio spectrum and the radial extent of minihalos is that it relies on a process, namely the amplification of the magnetic field by sloshing, that is strongly evidenced by previous theoretical and simulation work, and is independent of the details of the underlying CRp spectrum and spatial distribution. For this reason, we believe it deserves close examination.

In this work, we use the same cluster setup as in Z13 to determine if the amplification of the magnetic field strength by the sloshing motions is sufficient {\it by itself} to produce the two distinct characteristics mentioned above, e.g., emission confined to the cold front region with a steep spectrum, steeper than that expected from the canonical CRp slope under stationary conditions. We follow the evolution of CRe spectra associated with passive tracer particles which are advected with the
flow of gas, determining the injection into and radiative losses on the
spectrum along each trajectory. In this way we can determine how
deviations from a steady-state configuration affect the surface
brightness of the minihalo emission as well as its spectral shape. 

This paper is organized as follows. In Section 2 we
describe our method for evolving the CRe spectra and our assumptions
regarding their injection by hadronic processes. In Section 3 we detail the results of our simulations. In Section 4 we discuss the implications of this work, and in
Section 5 we make our conclusions. Thoughout this work
we assume a flat $\Lambda$CDM cosmology with $h$ = 0.7 and
$\Omega_{\rm m}$ = 0.3. 

\section{Method\label{sec:method}}

\subsection{MHD Simulation\label{sec:MHD}}

For this work, we used the magnetohydrodynamic (MHD) simulation of a sloshing cluster core from Z13. We briefly outline the numerical methods and initial setup of this simulation, and refer the reader to ZML11 and Z13 for further algorithmic details. 

We performed our MHD simulation of gas sloshing using the \code{FLASH} code \citep{dub09}. \code{FLASH} solves the equations of ideal MHD using a directionally unsplit staggered mesh algorithm \citep[USM;][]{lee09}. For simulating the gravitational potential of the two clusters, we use the ``rigid body'' approach described in \citet{rod12}. Our initial conditions consist of a massive ($M \sim 10^{15} M_\odot, T \sim 10$~keV) cool-core galaxy cluster, and a smaller ($M \sim 2 \times 10^{14} M_\odot$) gasless subcluster, separated at an initial distance $d$ = 3~Mpc, an initial impact parameter $b$ = 500~kpc, and placed on a bound mutual orbit. Gas sloshing is initiated in the massive cluster by the gravitational encounter with the subcluster. 

Our simulation employs adaptive mesh refinement (AMR). The size of the computational domain is $L$ = 2~Mpc, with a finest cell size of $\Delta{x}$ = 1~kpc (a discussion of the impact of the finite resolution of our simulation on the amplification of the magnetic field strength is given in the Appendix). The maximum AMR resolution covers a spherical region of $r \sim 300$~kpc, centered on the DM peak of the main cluster, which encompasses all of the phenomena of interest in this study. We set up the initial magnetic field on the grid with a Gaussian random field with a Kolmogorov spectrum. The magnetic field strength has a radial profile that declines as the square root of the thermal pressure (e.g., a constant $\beta = p_{\rm th}/p_B$ = 100). As detailed in Z13, we allowed this
configuration to relax for several Gyr. The resulting magnetic field profile from this
relaxation peaks at roughly $B \sim 12~\mu$G at the cluster center,
decreases to $\sim$5~$\mu$G at the cooling radius, and drops below
$B_{\rm CMB}$ (the redshift-dependent equivalent magnetic field strength of the CMB energy, $\approx 3.24~(1+z)^2~\mu{G}$) at $r \sim 100$~kpc (cf. Figure 1 of ZML11). 

Finally, our simulation includes passive tracer particles flowing with the ICM, which carry with them CRe spectra which we evolve according to the method in Section \ref{sec:method_spec}. Tracer particles are stored at intervals of 10~Myr and from these snapshots the individual particle trajectories are extracted. The entire simulation contains approximately 10 million passive tracer particles initially distributed with their number density proportional to the local gas density. The local conditions of the gas at any time, such as the density of thermal particles and the magnetic field, can be determined by linearly interpolating between their values at the epochs corresponding to the datasets.

\subsection{Method for Evolution of the Relativistic Electron
  Spectrum\label{sec:method_spec}}

We model the CRe component as time-dependent spectra $N(p,t)$ of the CRe momentum $p$ along the tracer particle trajectories (in post-processing). In this approximation, the CRe are assumed to passively advect with the tracer particles once they are injected. Along each tracer particle trajectory, we employ a kinetic equation for $N(p,t)$ (equivalent to a Fokker-Planck equation without the diffusive terms):
\begin{equation}
\frac{\partial{N(p,t)}}{\partial{t}} + \frac{\partial}{\partial{p}}\left[N(p,t)\frac{dp}{dt}\right] = Q(p,t)
\end{equation}
where $dp/dt$ is the sum of the standard radiative and Coulomb losses: 
\begin{eqnarray}
\frac{dp}{dt} &=& \left.\frac{dp}{dt}\right|_{\rm rad} + \left.\frac{dp}{dt}\right|_{\rm coul} \\
\left.\frac{dp}{dt}\right|_{\rm rad} &=& -4.8 \times 10^{-4}p^2\left[\left(\frac{B_{\rm \mu{G}}}{3.25}\right)^2+(1+z)^4\right] \\
\left.\frac{dp}{dt}\right|_{\rm coul} &=& -3.3 \times 10^{-29}n_{\rm th}\left[1+\ln\left(\frac{\gamma/n_{\rm th}}{75}\right)\right],
\end{eqnarray}
where the redshift is evolved according to our assumed cosmology. We assign $z = 0$ to the epoch $t$ = 5~Gyr of the simulation, in order to reproduce some of the observed nearby clusters exhibiting cold fronts in their cores. Our calculations do not include reacceleration of CRe by shocks or turbulence, or changes in energy due to adiabatic compression and expansion of the gas.

To solve this kinetic equation we logarithmically sampled the CRe
spectrum $N(p)$ 100 times in the range of $p/m_ec \in [10,10^5]$, and
evolved it on this momentum grid forward in time using a finite
difference method. At each timestep, we separate the calculation into two steps,
``advection'' and ``source term'', using operator splitting. In the
``advection'' step, we compute the change in the spectrum in each
momentum cell due to the change in the flux of electrons
$\partial/\partial{p}[Ndp/dt]$ from the adjacent cells on the momentum
grid. To ensure stability, we applied a minmod slope limiter
to this flux at each cell boundary. Following the ``advection'' step, we apply the ``source term'' operation $Q(p,t)$ by computing the change in the spectrum at each momentum cell using the time-interpolated value of the source term. 

Finally, at the lower momentum boundary we impose outflow
boundary conditions, and at the higher momentum boundary we extend the
spectrum into the boundary cells with a constant logarithmic slope, a
sensible choice since at such high energies in a steady-state the spectrum is dominated
by the competition between injection and radiative losses and the
spectrum is essentially a pure power law with index $\alpha_p+1$.

We verify our model for the evolution of the spectrum by confirming that it can reproduce analytic predictions for the evolution of the CRe spectrum under simplified conditions (e.g., radiative and Coulomb losses without injection, and a steady-state balance between injection and losses; see the left panel of Figure \ref{fig:CRe_spec}.)

\subsection{Modeling the Injection of Secondary Electrons\label{sec:method_injection}}

The injection rate is determined by the rate of interactions of CRp
with the thermal population. We assume that the CRp momentum ($p_p =
\gamma_p\beta_p{m_pc}$) spectrum can be approximated by a single
power-law, $N_p(p_p) = C_pp_p^{-\alpha_p}$, where $C_p$ is a
normalization constant and $\alpha_p = 2.3$, a value within the range ($\approx$~2.0-2.5)
implied by various mechanisms capable of injecting relativistic CRp
into the ICM \citep[e.g., structure formation shocks, injection by
radio galaxies, supernova remnants;][]{pfr04,kes10a}.

For consistency with previous literature, we follow in part the derivation of the electron source function of \citet{pfr08}.\footnote{Note that in our derivation the symbol $p$
  corresponds to the particle momentum but in their derivation it
  corresponds to the momentum nomalized by $mc$.} The pion production
spectrum is   
\begin{equation}
s_\pi(p_\pi, p_p) = cn_{\rm th}\xi(p_p)\sigma_{\rm pp}^\pi(p_p)\delta_D(p_\pi-\langle{p_\pi}\rangle)
\end{equation}
where $\sigma_{\rm pp}^\pi$ is the inelastic p-p cross section and
$\langle{p_\pi}\rangle$ is the average momentum of a single produced
pion, and we have employed the delta-function approximation. We may integrate this source function with respect to the CRp momentum distribution to obtain
\begin{equation}
s_{\pi^\pm}(p_\pi) = \frac{2}{3}q_\pi(p_\pi) = \frac{2}{3}\displaystyle\int_{-\infty}^{\infty}f(p_p)s_\pi(p_\pi,p_p)dp_p
\end{equation}
where $\langle{p_\pi^\pm}\rangle = p_p/4$. 

Using the transformation law for distribution functions and using the
mean value of the CRe momentum in the relativistic limit, we
approximate the CRe momentum source function by 
\begin{eqnarray}
Q(p)dp &=& s_\pi^\pm(p_\pi)\frac{dp_\pi}{dp}dp \\
            &=& \frac{4}{3}16^{1-\alpha_p}cn_{\rm th}C_p\sigma_{\rm pp}(16p)p^{-\alpha_p}dp
\end{eqnarray}
assuming that the mean momentum of the produced
secondary electrons in the laboratory frame is given
by $\langle{p_e}\rangle = \frac{1}{4}\langle{p_\pi^\pm}\rangle$.
Previous works \citep[see, e.g.][]{dol00,pfr04,pfr08} have commonly
approximated the proton-proton cross section by an energy-averaged
constant value, which results in a pure power-law form for the source
function, and hence the steady-state electron CRe is also a power-law
with spectral index $\alpha_e = \alpha_p + 1$, and the synchrotron
spectral index is $\alpha = (\alpha_e-1)/2 = \alpha_p/2$. In this work we also account for the weak energy dependence of the cross
section on the proton energy found by \citet{kel06}:
\begin{equation}
\sigma_{\rm pp}(E_p) = (34.3+1.88L+0.25L^2) \times
\left[1-\left(\frac{E_{\rm th}}{E_p}\right)\right]^2
\label{eqn:sigma_pp}
{\rm mb} 
\end{equation}
where $L = \ln(E_p/{\rm 1~TeV})$ and $E_{\rm th} = m_p + 2m_\pi
+m_\pi^2/2m_p = 1.22 \times 10^{-3}~{\rm TeV}$ is the threshold energy of
production of $\pi^0$ mesons.

Our inclusion of a weak energy dependence of $\sigma_{\rm
  pp}$ results in a steady-state CRe spectrum with a slope that is
slightly shallower than $\alpha_p+1$. From numerical tests assuming a
steady-state spectrum of CRe we have determined that the observed
spectral index for magnetic fields of $B \sim 0.1-10~\mu$G will be
roughly within the range of $\alpha \sim 1.09-1.17$, with a typical
value of $\alpha \approx 1.13$ (see Figure
\ref{fig:alpha_histogram}),\footnote{Unless otherwise noted we fit for $\alpha$ between 327 and 1420~MHz.} a very minor deviation from the value of 1.15
expected by treating $\sigma_{\rm pp}$ as constant.

\subsection{Modeling the Cosmic-Ray Protons\label{sec:method_crp}}

\begin{figure*}
\begin{center}
\includegraphics[width=0.32\linewidth]{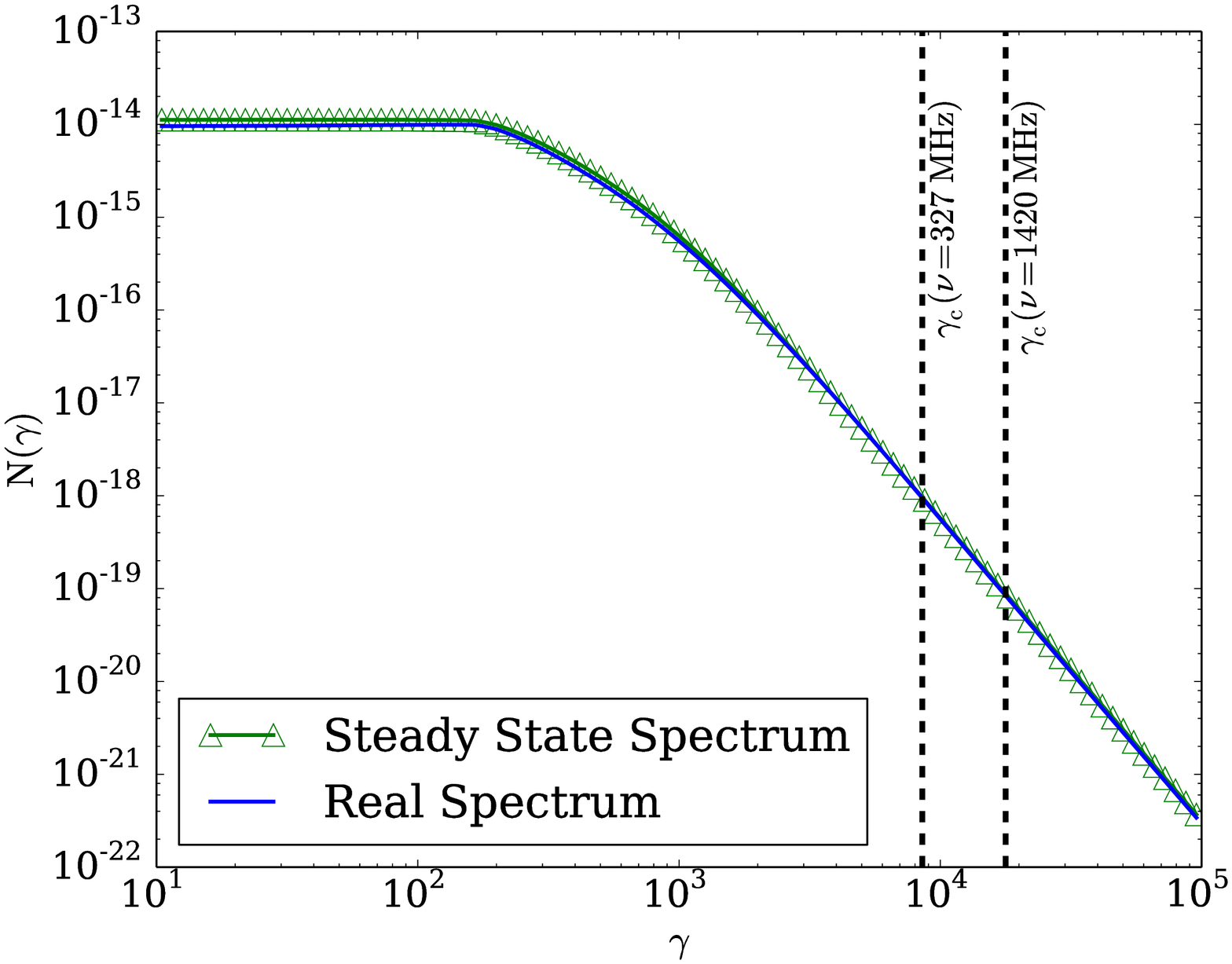}
\includegraphics[width=0.32\linewidth]{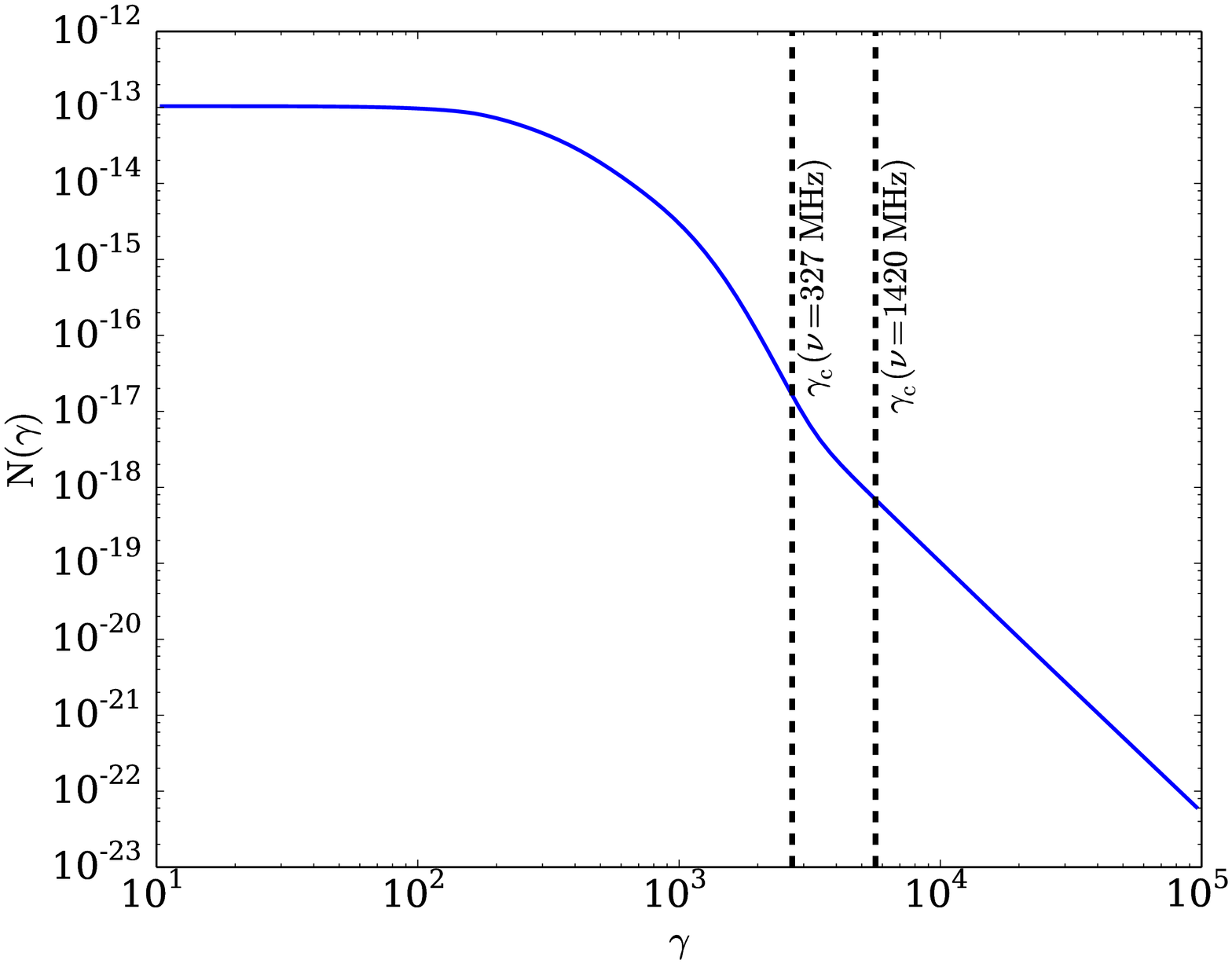}
\includegraphics[width=0.32\linewidth]{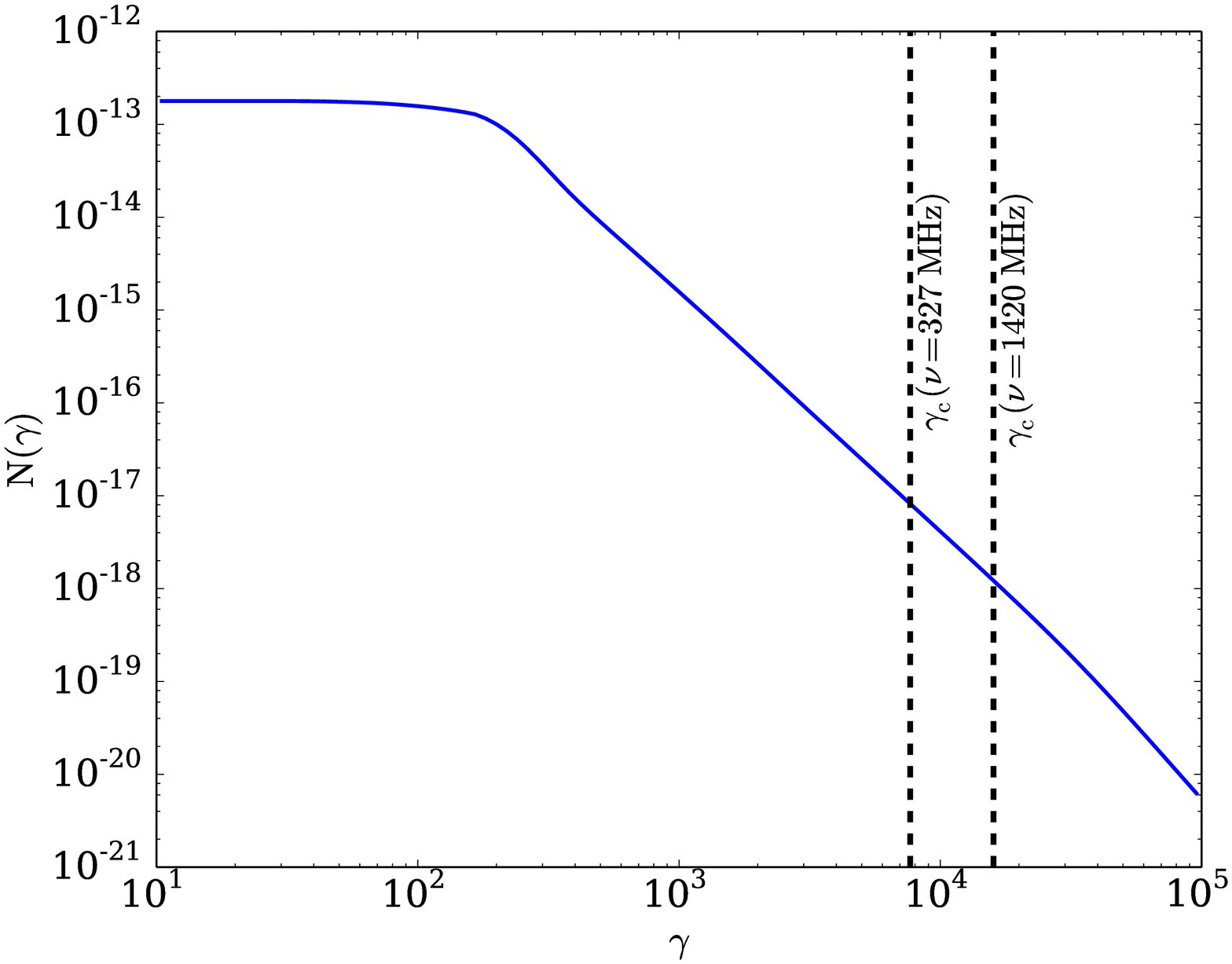}
\caption{CRe spectra under various conditions. Left: A typical CRe
  spectrum for a particle close to the steady-state condition. The
  blue line indicates the real particle spectrum,and the green line
  indicates the steady-state spectrum for the same conditions. Center:
  A typical CRe spectrum for a particle with a steep synchrotron
  spectrum. Right: A typical CRe spectrum for a particle with a shallow
  synchrotron spectrum. In each panel, the vertical dashed lines
  indicate the characteristic CRe energies corresponding to the
  frequencies of 327 and 1420~MHz.\label{fig:CRe_spec}}
\end{center}
\end{figure*}

The CRe in our simulations are produced solely by hadronic interactions between CRp and thermal protons. CRp are believed to be injected in the ICM by a variety of processes, including shocks from supernovae, large-scale structure formation, and active galactic nuclei \citep{pfr04,kes10a}. 

In particular, diffusion (along the field lines) is an energy-dependent process that depends on the spectrum of MHD waves and on the physics of the interplay/scatter between CRs and these waves. The time necessary for CRs to diffuse
over distances $L$ is $\tau_{\rm diff} \sim 1/4 L^2/D$, where $D$ is
the spatial diffusion coefficient. This process is expected to affect
our results if CRs diffuse on scales of about 100~kpc within a time
frame of $\sim$~1~Gyr, which would imply a diffusion coefficient $D
\geq 10^{30}$cm$^2$/s that is very large, although it cannot be ruled out. 

Sloshing can generate substantial turbulence in the core region. Under these conditions, CRs are expected to be transported as passive
scalars by the turbulent gas-flow \citep[][]{bru14}. On scales
similar to that of the turbulence or larger, this induces a diffusive process
that is energy independent with a diffusion coefficient $D \sim V_o l_o$, where $V_o$
and $l_o$ are the velocities and scales of the largest turbulent
eddies. According to numerical simulations of sloshing motions in
galaxy clusters, $D < 10^{29}$~cm$^2$/s \citep{vaz12b}, implying
typical transport-diffusion scales of $\sim 10-50$~kpc on a timescale of
1~Gyr for CRp.

Previous works \citep{min01a,min01b,pfr07,pfr08,vaz12a,vaz14} have studied the generation, advection and energy evolution of CRp in the cosmological context. Our focus in this work is on the effects produced by subsonic sloshing motions in a idealized, relatively relaxed cluster. Therefore, we have chosen to model the evolution of the CRe spectrum under the simplifying assumption that the CRp energy density is proportional to the thermal energy, $\epsilon_{\rm CRp} = X_{\rm CRp}\epsilon_{\rm th}$, e.g., we assume passive advection of the CRp population with the ICM and ignore the propagation effects and energy loss processes mentioned above (we briefly discuss the limitations of this approach in Section \ref{sec:limitations}.) We choose a reference value of $X_{\rm CRp} = 0.02$, at the upper limit placed on the energy density of cosmic-ray protons by the Fermi, HESS, and MAGIC non-detections of $\gamma$-ray emission from neutral pions, which are produced in the same hadronic interactions and subsequently decay
\citep[][]{ack10,ale12,fer13,pro13}. 

\section{Results\label{sec:results}}

\begin{figure*}
\begin{center}
\plottwo{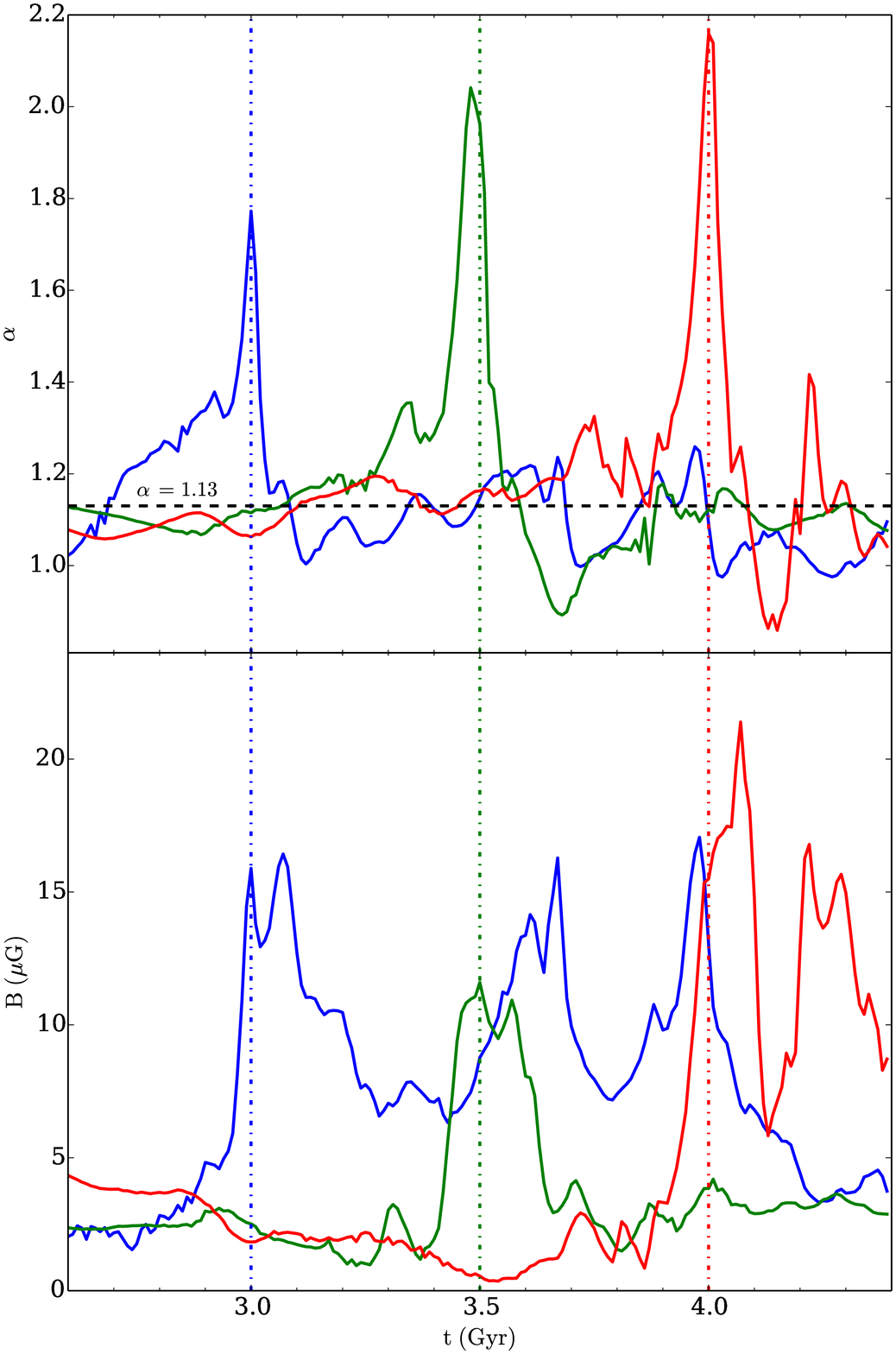}{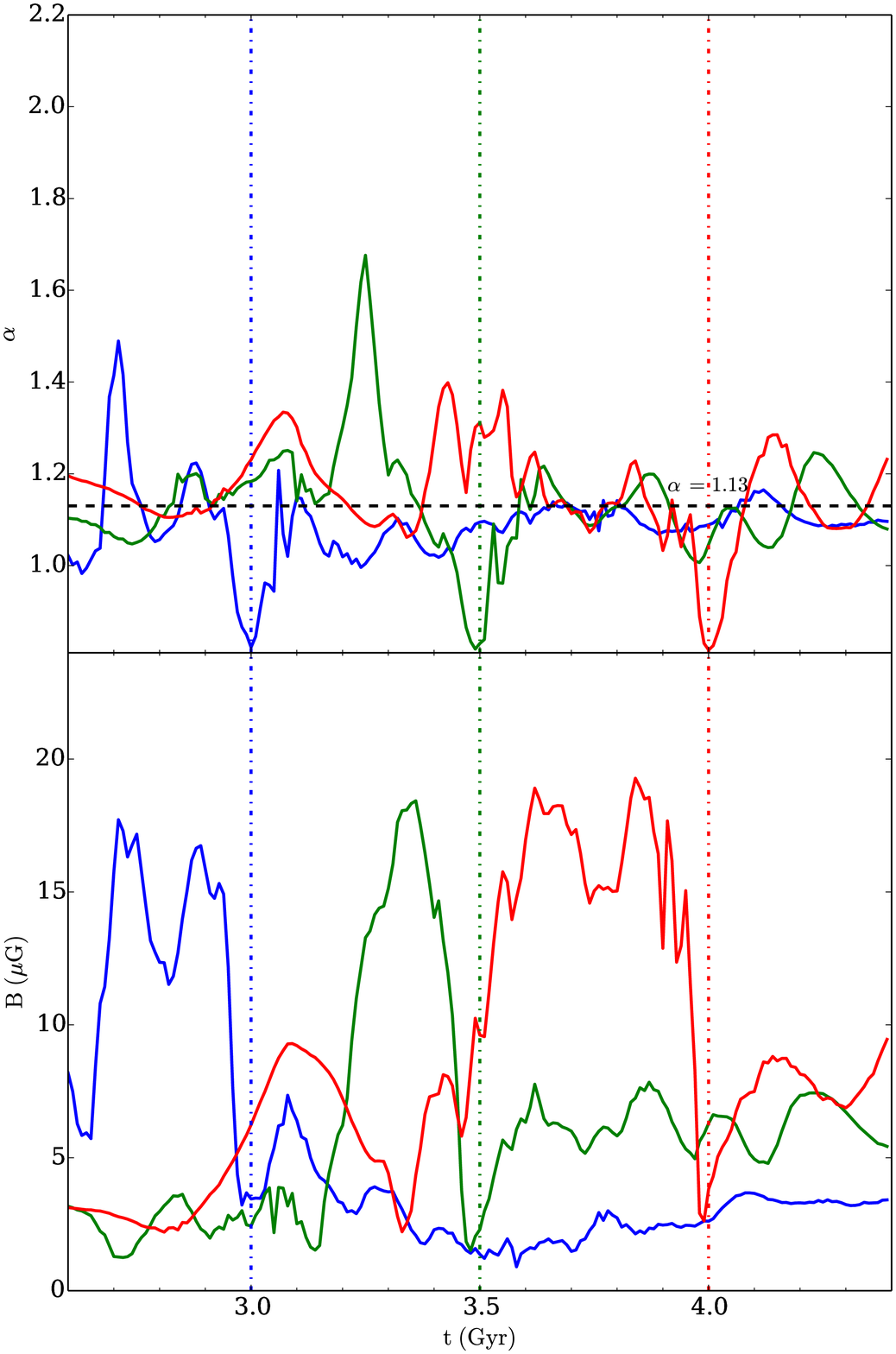}
\caption{The evolution of the spectral index $\alpha$ and the magnetic
  field strength over time for selected particles. Left panel:
  Particles with steep spectra at a given epoch. Right panel:
  Particles with shallow spectra at a given epoch. Colored vertical
  dot-dashed lines indicate the epochs at which the extreme spectra
  occur for each particle. The horizontal black line indicates the
  approximate steady-state spectral index.\label{fig:alpha_evolution}}
\end{center}
\end{figure*}

\subsection{Tracer Particles Evolution\label{sec:tracer_particles}}

To get a sense for how deviations from steady-state conditions affect
the resulting electron spectrum of the particles, we will first
examine the properties of the tracer particles. 

The left panel of Figure \ref{fig:CRe_spec} shows a typical example of a tracer
particle with a steady-state CRe spectrum at the epoch $t = 3.5$~Gyr,
approximately 2~Gyr past the core passage, a moment at which there is
strong magnetic field amplification due to sloshing. The magnetic
field strength for this particular particle, however, is not changing
rapidly at this epoch. The black dashed lines indicate the values of
$\gamma$ that roughly correspond to the frequencies of 327~MHz and
1420~MHz at $B \approx 1~\mu$G. At low energies, where no injection
occurs (due to the energy threshold for pion production, cf. Equation \ref{eqn:sigma_pp})
and Coulomb losses dominate, the spectrum is relatively
flat, whereas at higher energies it transitions to a power-law shape
with $\alpha_e \approx 1.13$, where it is dominated by the tradeoff
between injection and radiative losses. For comparison, the
steady-state spectrum for the same gas and magnetic field conditions
is also plotted in the left panel of Figure \ref{fig:CRe_spec}. The near equivalence of these two spectra indicates that the magnetic field strength
and local density of particles are varying smoothly enough that the
steady-state assumption is adequate. 

The center panel of Figure \ref{fig:CRe_spec} shows a typical example of a CRe spectrum
of a tracer particle with a steep spectrum ($\alpha \approx 2$) at the
epoch $t = 3.5$~Gyr. The magnetic field of this particle is undergoing significant
amplification at this epoch. The black dashed lines indicate the values
of $\gamma$ that roughly correspond to the frequencies of 327~MHz and
1420~MHz at $B \approx 12~\mu$G. This spectrum, in contrast to the
example spectrum that is close to the steady-state case, has a steep
transition segment between the flat segment and the high-energy,
$\alpha \approx 1.13$ segment which is determined by the tradeoff
between injection and radiative losses. 

The right panel of Figure \ref{fig:CRe_spec} shows a typical example of a CRe
spectrum of a tracer particle with a shallow spectrum ($\alpha \approx
0.8$) at the epoch $t = 3.5$~Gyr. The magnetic field strength of this
particle is decreasing rapidly at this epoch. The black dashed lines indicate the values
of $\gamma$ that roughly correspond to the frequencies of 327~MHz and
1420~MHz at $B \approx 2.3~\mu$G. Over most of the range of particle energies, the
spectral index is nearly constant, except at very high energies where
it steepens due to the $\propto p^2$ dependence of the radiative
losses. 

It is instructive to examine the behavior of the spectral index of
individual tracer particles to see how they change with time, and how
this is related to the change with time of the magnetic field
strength. Figure \ref{fig:alpha_evolution} shows the change in the
spectral index and the magnetic field strength for three different
particles each which at some point have steep and shallow spectra,
respectively. The particles we have examined in Figure
\ref{fig:CRe_spec} are indicated by the green curves. These curves
demonstrate that while most of the time the spectral index is fairly
constant and close to the steady-state value, that when the magnetic
field increases rapidly it steepens, and when there is a rapid
decrease it shallows. Although these effects are clear and
theoretically expected \citep{kes10b} for particular tracers, we will
show later (within this section and Section \ref{sec:spec_index}) that
they cannot be representative. 

\begin{figure}
\begin{center}
\plotone{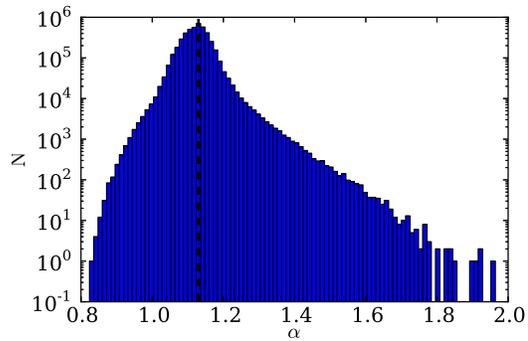}
\caption{Binned histogram of tracer particle spectral indices at $t$
  = 3.5~Gyr. The dashed line indicates $\alpha$ = 1.13.\label{fig:alpha_histogram}}
\end{center}
\end{figure}

Figure \ref{fig:alpha_histogram} shows a binned histogram of the radio spectral index
$\alpha$ for all of the tracer particles at the epoch $t$ = 3.5~Gyr. It has a narrow peak at $\alpha \approx 1.13$ and a small skew toward steeper spectral indices. The spectra of particles gets as flat as $\alpha \approx 0.8$ and as steep as $\alpha \sim 2$, but the number of these particles is dwarfed by the number of particles with spectra near the
steady-state slope.

\begin{figure}
\begin{center}
\plotone{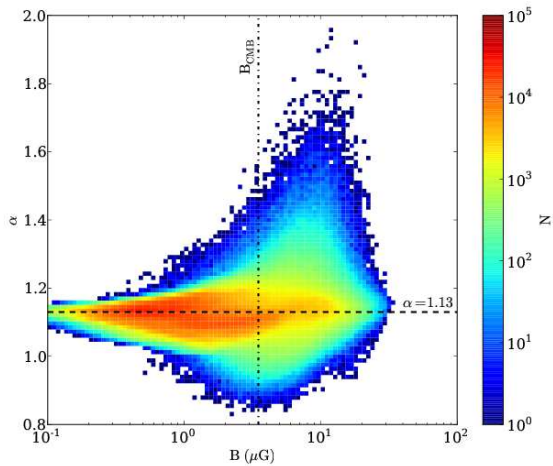}
\caption{Phase plot of the synchrotron spectral index versus the
  magnetic field strength at the epoch $t$ = 3.5~Gyr. The horizontal dashed
  line indicates $\alpha = 1.13$. The vertical dot-dashed line
  indicates where $B = B_{\rm CMB}$.\label{fig:B_vs_alpha}}
\end{center}
\end{figure}

To get a sense of how the spectral index depends on the magnetic
field, Figure \ref{fig:B_vs_alpha} shows a phase plot of the radio spectral index versus
the magnetic field strength for all of the particles in the
simulation at the epoch $t$ = 3.5~Gyr. At low magnetic field strengths, the fitted spectral index is very close to the steady-state expectation for nearly all tracer
particles. The lack of steep spectral indices at these magnetic field
strengths is readily understood; in this regime $B \ll B_{\rm CMB}$. The fact that we also do not observe many particles to have shallower spectra at these magnetic field strengths is due to the fact that at lower magnetic field strengths the observed
synchrotron frequencies are sampling higher CRe energies, which will
be within the part of the spectrum that is determined essentially
completely by the tradeoff between radiative losses and injection.

As we go to higher magnetic field strengths, we see a broader distribution of
spectral indices. Some of these spectra are somewhat flatter than the
steady-state expectation, which correspond to particles where the
magnetic field has been dropping rapidly (Figure \ref{fig:alpha_evolution}). We find that
the number of these particles increases up to $B \approx B_{\rm CMB}$
and then drops off, due to the increasing importance of the
synchrotron over the inverse-Compton losses. Similarly, beginning
around $B \approx B_{\rm CMB}$, the population of tracer
particles with very steep spectra increases, corresponding to
particles which are experiencing rapid increases in magnetic field
strength. Nevertheless, the number of tracer particles with nearly
steady-state spectra are dominant. 

\subsection{Radio Intensity Maps}

\begin{figure*}
\begin{center}
\includegraphics[width=0.45\linewidth]{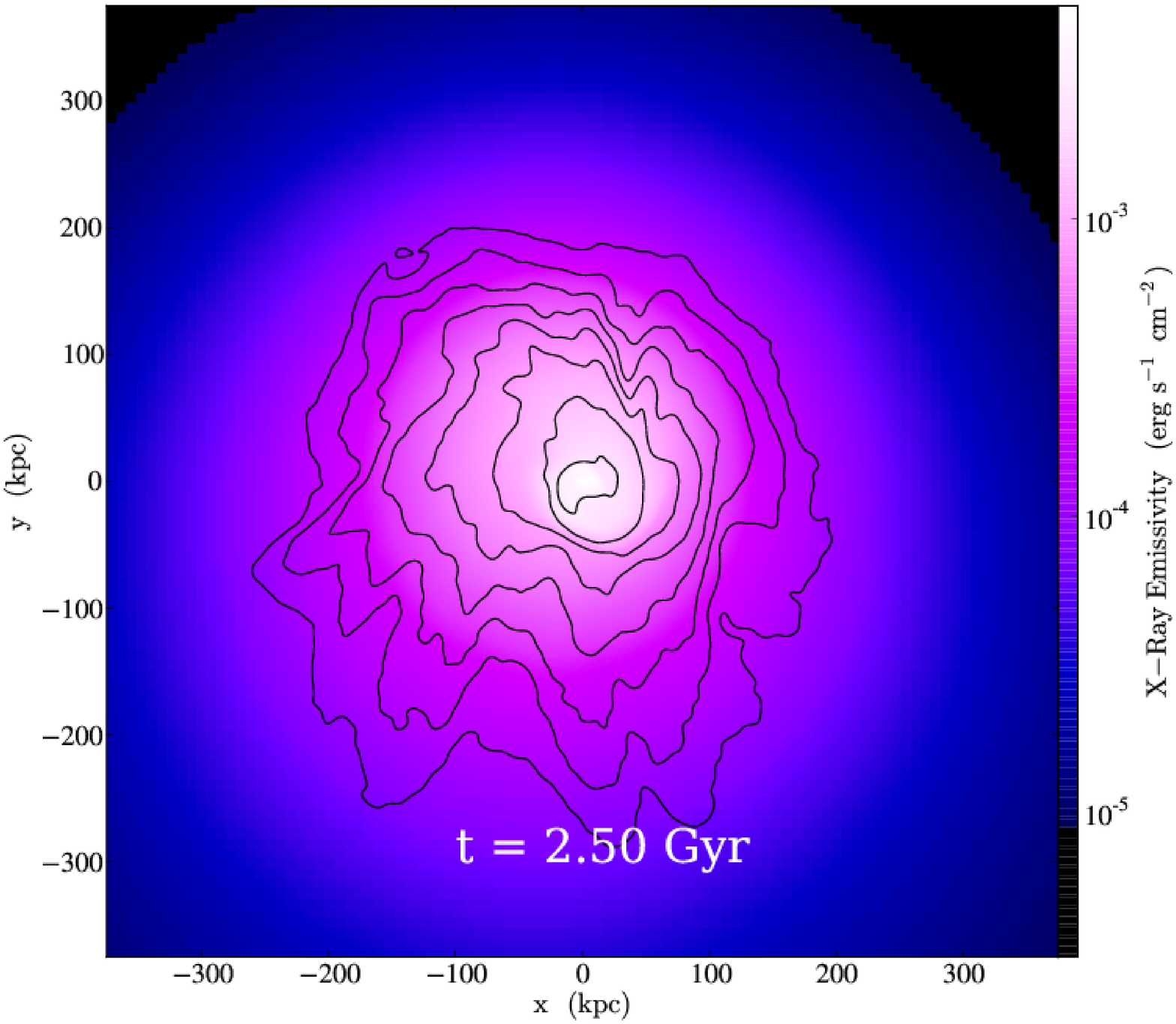}
\includegraphics[width=0.45\linewidth]{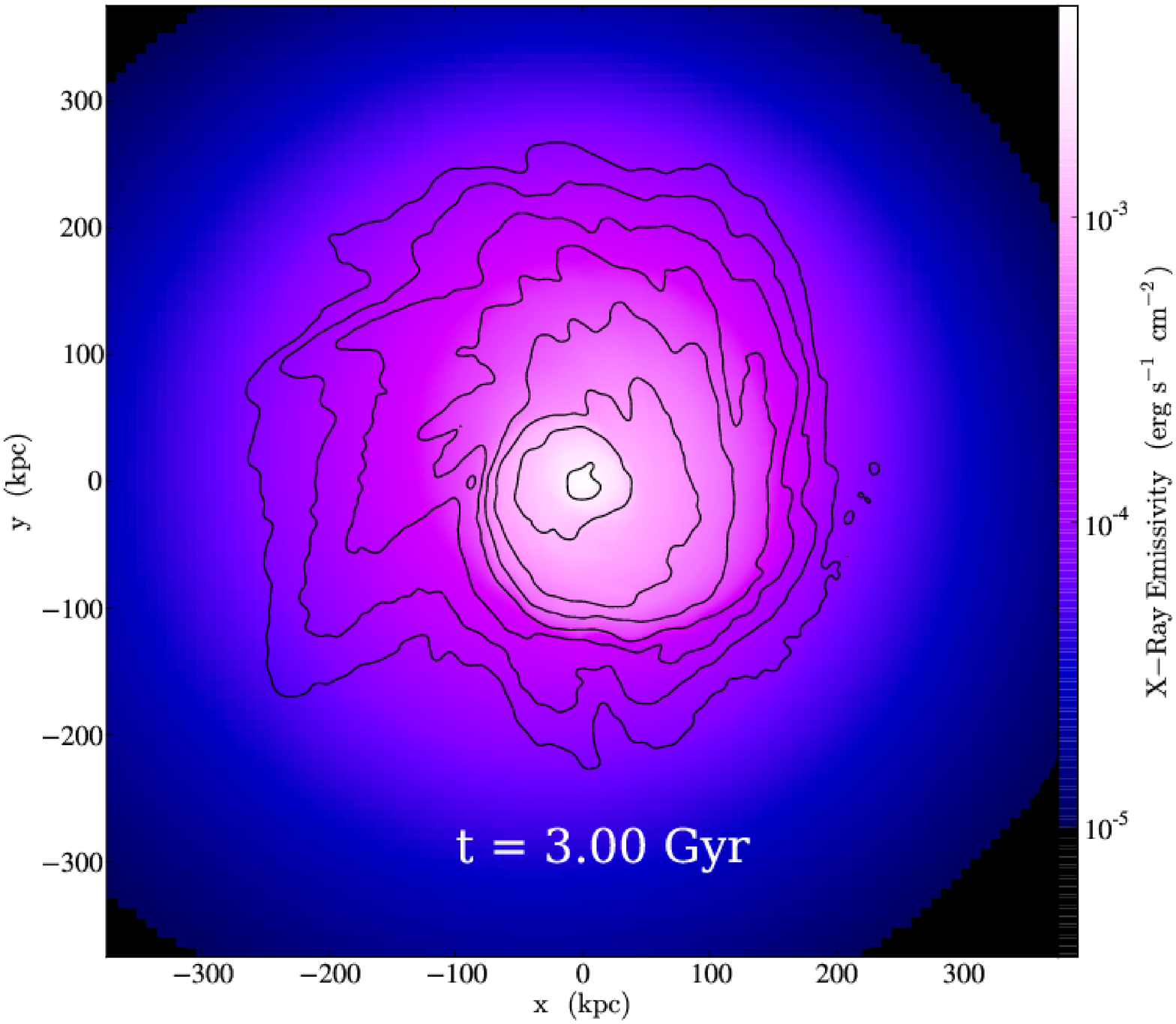}
\includegraphics[width=0.45\linewidth]{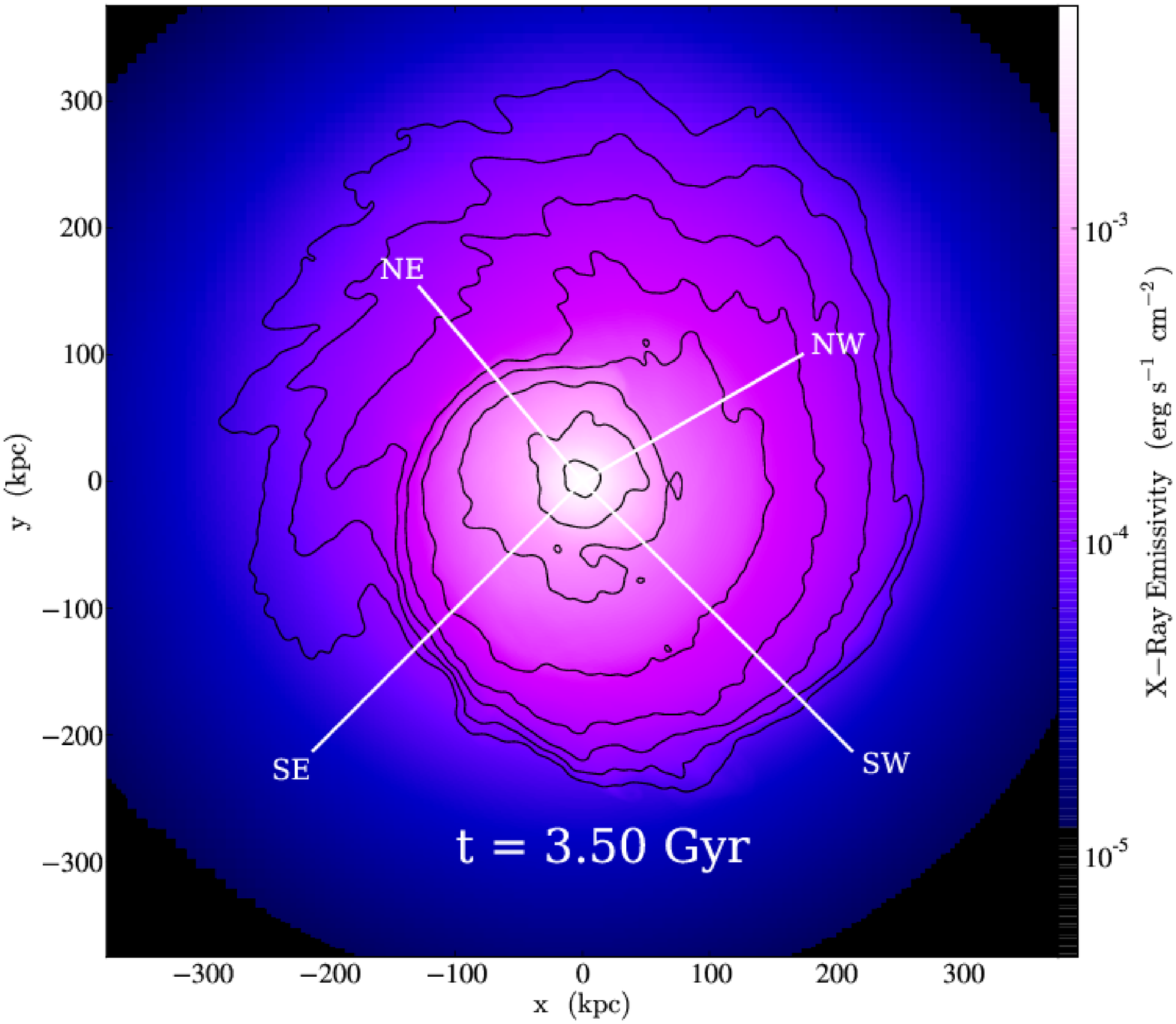}
\includegraphics[width=0.45\linewidth]{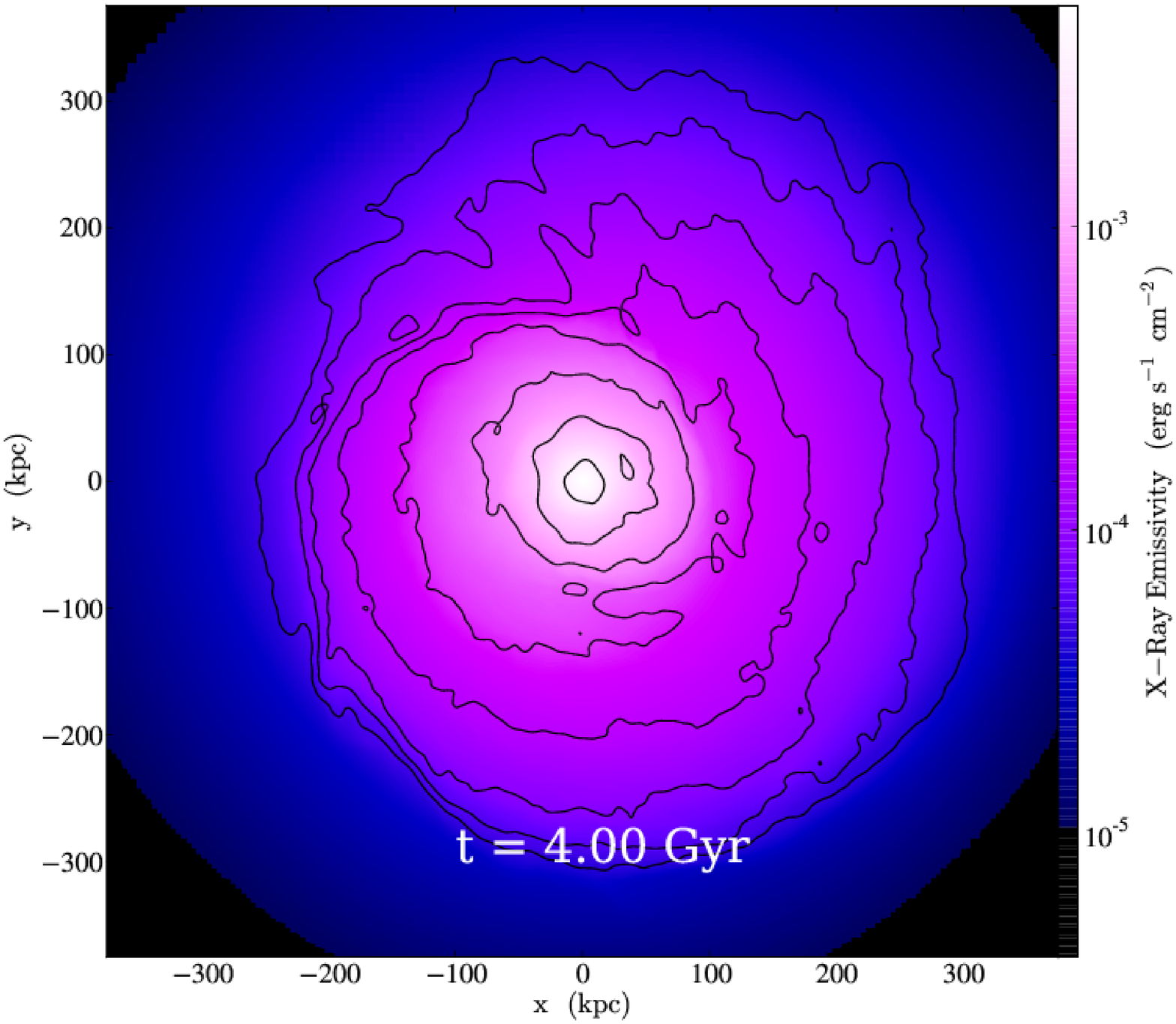}
\caption{Projected maps of X-ray emission with 327~MHz radio contours
  overlaid at four different epochs of the simulation. Contours begin
  at $10^{-4}$~mJy~arcsec$^{-2}$ and increase by a factor of 2. Each
  panel is 750~kpc on a side. Profiles in the bottom-left panel
  correspond to those in Figure \ref{fig:profiles}.\label{fig:327_maps}}
\end{center}
\end{figure*}

Figure \ref{fig:327_maps} shows maps of the X-ray emission of the
thermal gas with contours of radio surface brightness at the
frequency of 327 MHz overlaid for several different epochs of the
simulation where the CRp energy follows the thermal energy density. The resulting radio emission is diffuse and fills the cool-core region of the cluster, similar to observed minihalos. It is fairly extended, and largely reflects the shape of the X-ray emission, which is expected in this case since $j_\nu \propto n_{\rm th}^2$, which is the same dependence on density as the X-ray emission. In particular, the spiral shape of
the sloshing gas is readily seen in the radio contours.

\begin{figure*}
\begin{center}
\includegraphics[width=0.45\linewidth]{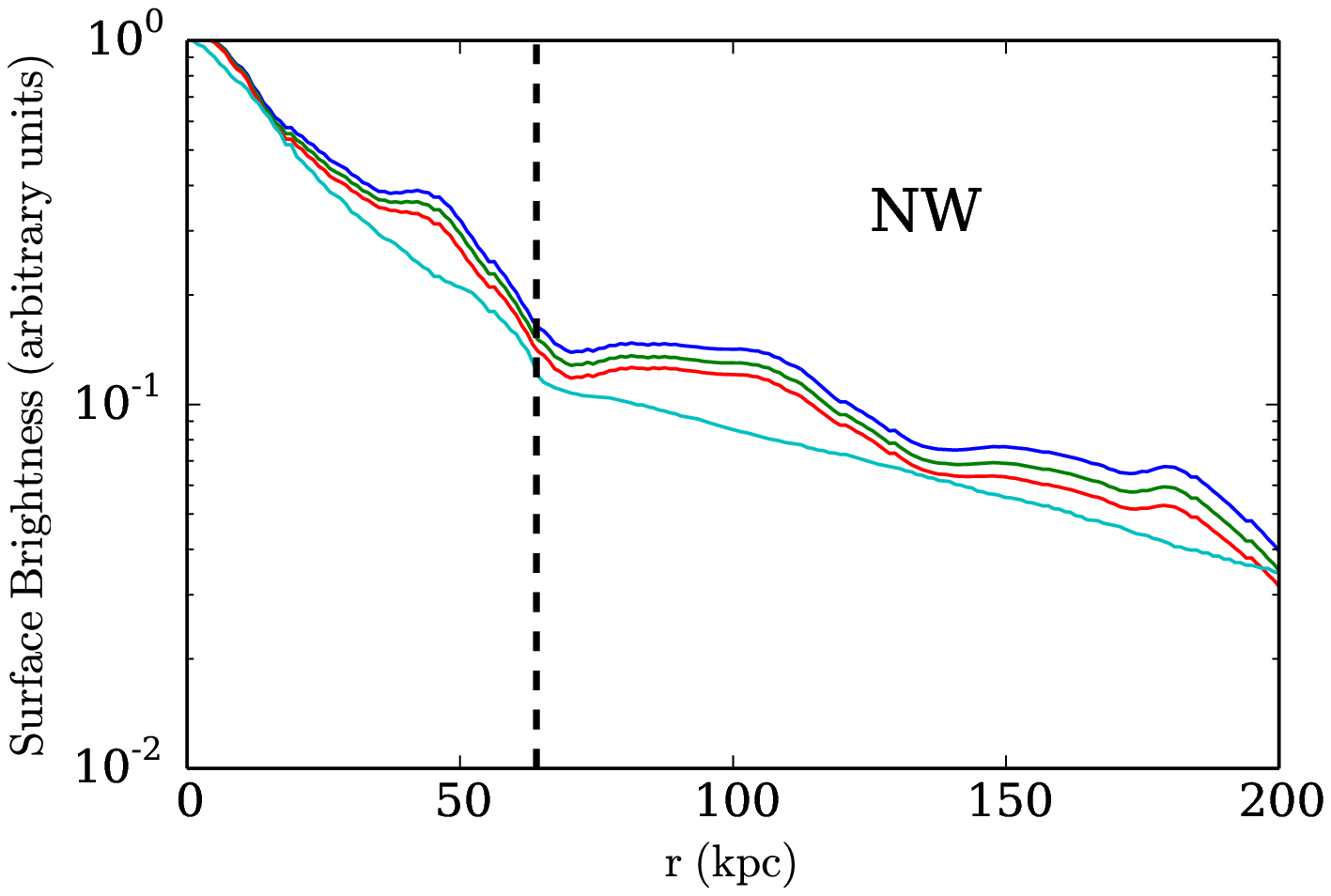}
\includegraphics[width=0.45\linewidth]{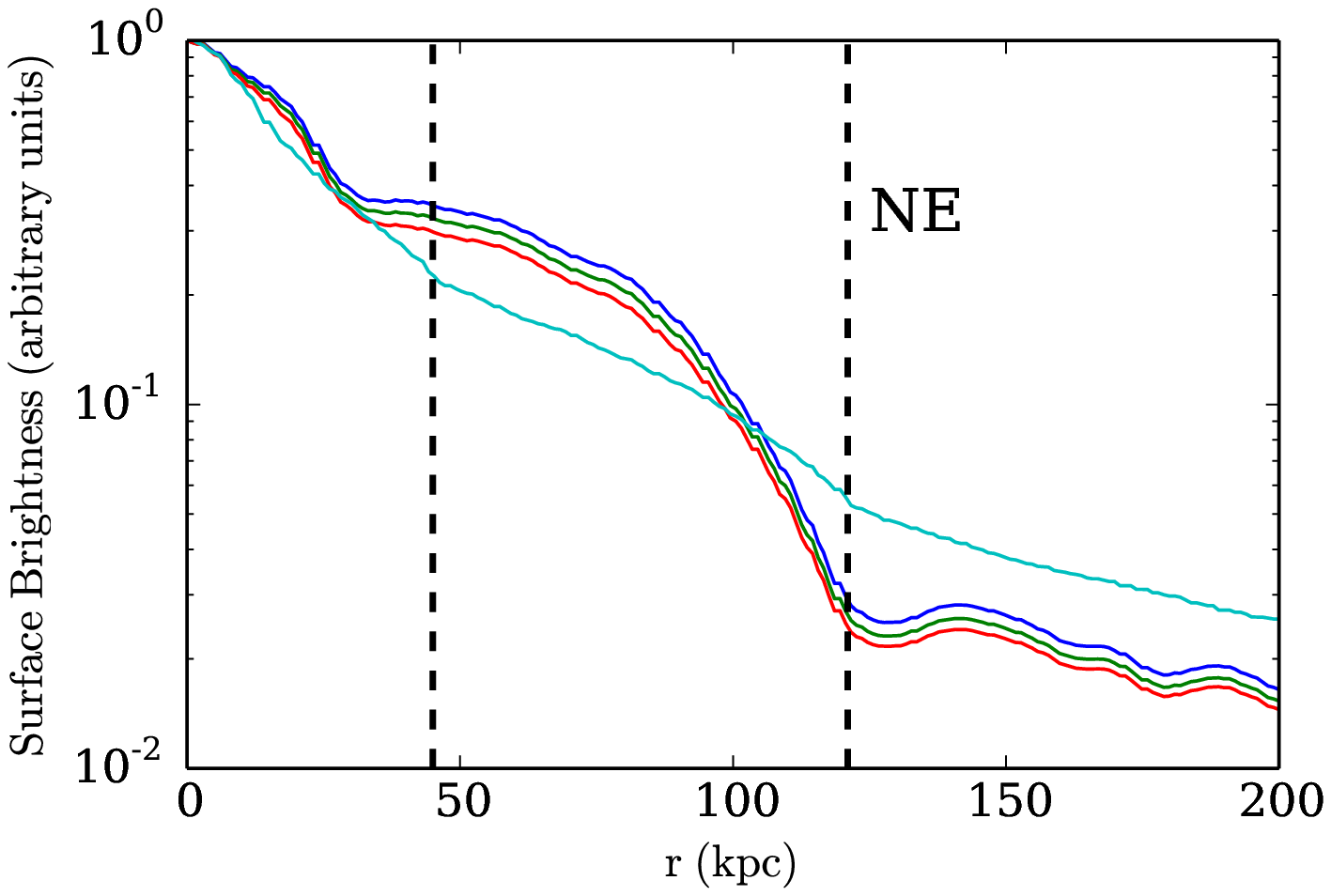}
\includegraphics[width=0.45\linewidth]{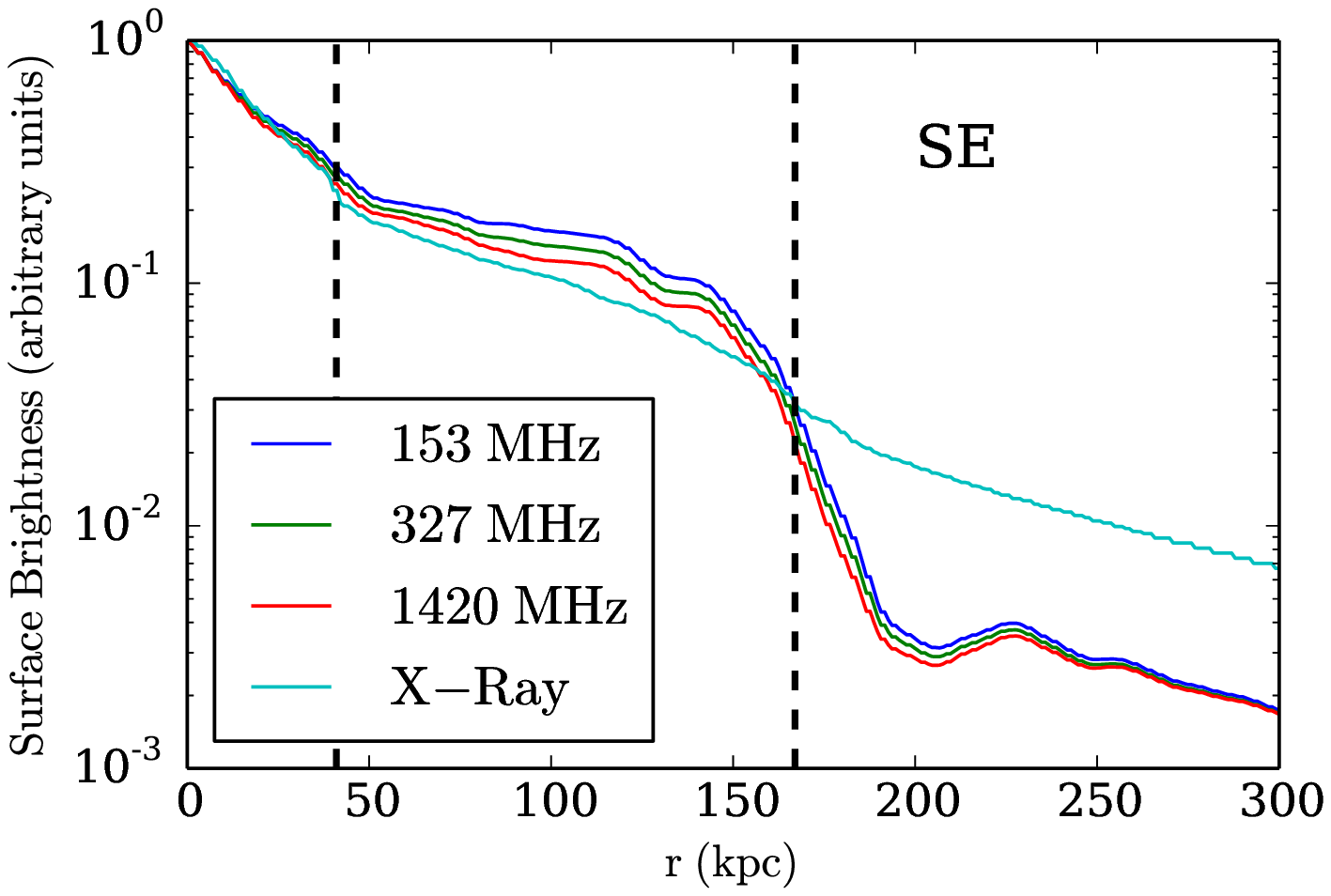}
\includegraphics[width=0.45\linewidth]{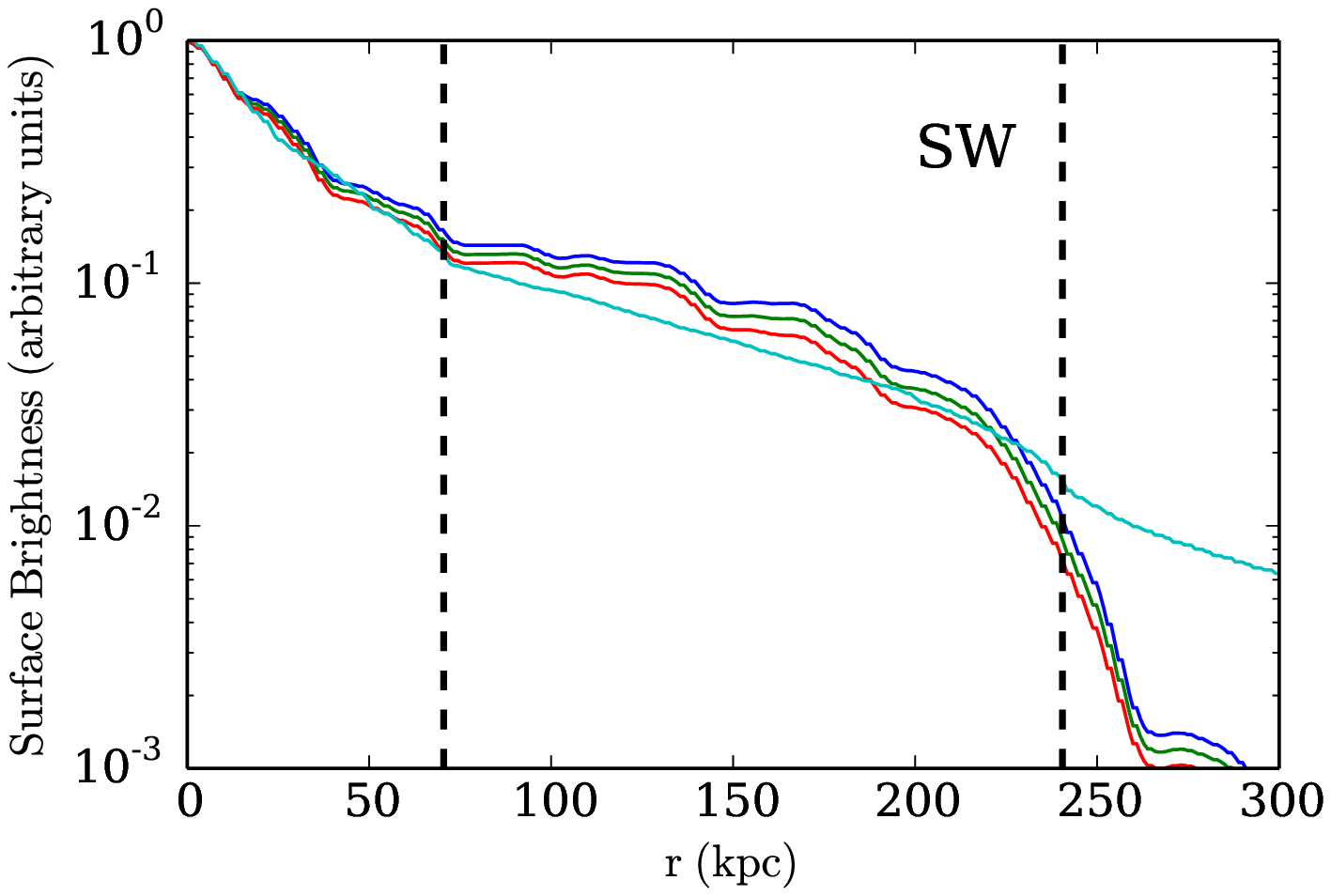}
\caption{Profiles of the X-ray and radio emission (at three different
  frequencies) at the epoch $t$ = 3.5~Gyr (see Figure
  \ref{fig:327_maps} for profile locations). The emission has been
  renormalized so that all profiles are unity at $r = 0$ for
  comparison purposes. The positions of the X-ray cold fronts are
  marked with black dashed lines.\label{fig:profiles}}
\end{center}
\end{figure*}

Figure \ref{fig:profiles} shows profiles of the radio surface
brightness at the frequencies of 153, 327, and 1420~MHz corresponding to the black profiles in Figure \ref{fig:327_maps} at the epoch $t$ = 3.5~Gyr. Additionally, we have
plotted the X-ray surface brightness along these profiles, and rescaled the overall
normalization of both so that their radial dependence may be closely
compared. Along the southern profiles, there are drops in radio
emission across the cold front surface of an order of magnitude over
$\sim$20~kpc. Along the NE profile, there is a similar drop, but
it is over a larger radial range, $\sim$40~kpc. Along the NW profile,
the radio emission drops off with radius at roughly the same rate as
the X-ray emission. This plot also demonstrates that the overall shape
of the radio emission is roughly independent of frequency along these
profiles. This is expected, since once non-stationary effects play a
minor role, the spectrum is a power law and the properties of the minihalo simply scale with frequency. Therefore, the main driver of the behavior of the minihalo profile is the strength of the magnetic field along this profile. Along most directions, ${B^2 \gg B_{\rm CMB}^2}$ below the front and ${B^2 \ll B_{\rm CMB}^2}$ above, with the exception of the NW profile, where the magnetic field does not decrease substantially across the front surface (cf. Figure \ref{fig:scaledB} and Section \ref{sec:radial_extent}).

\begin{figure*}
\begin{center}
\includegraphics[width=0.43\textwidth]{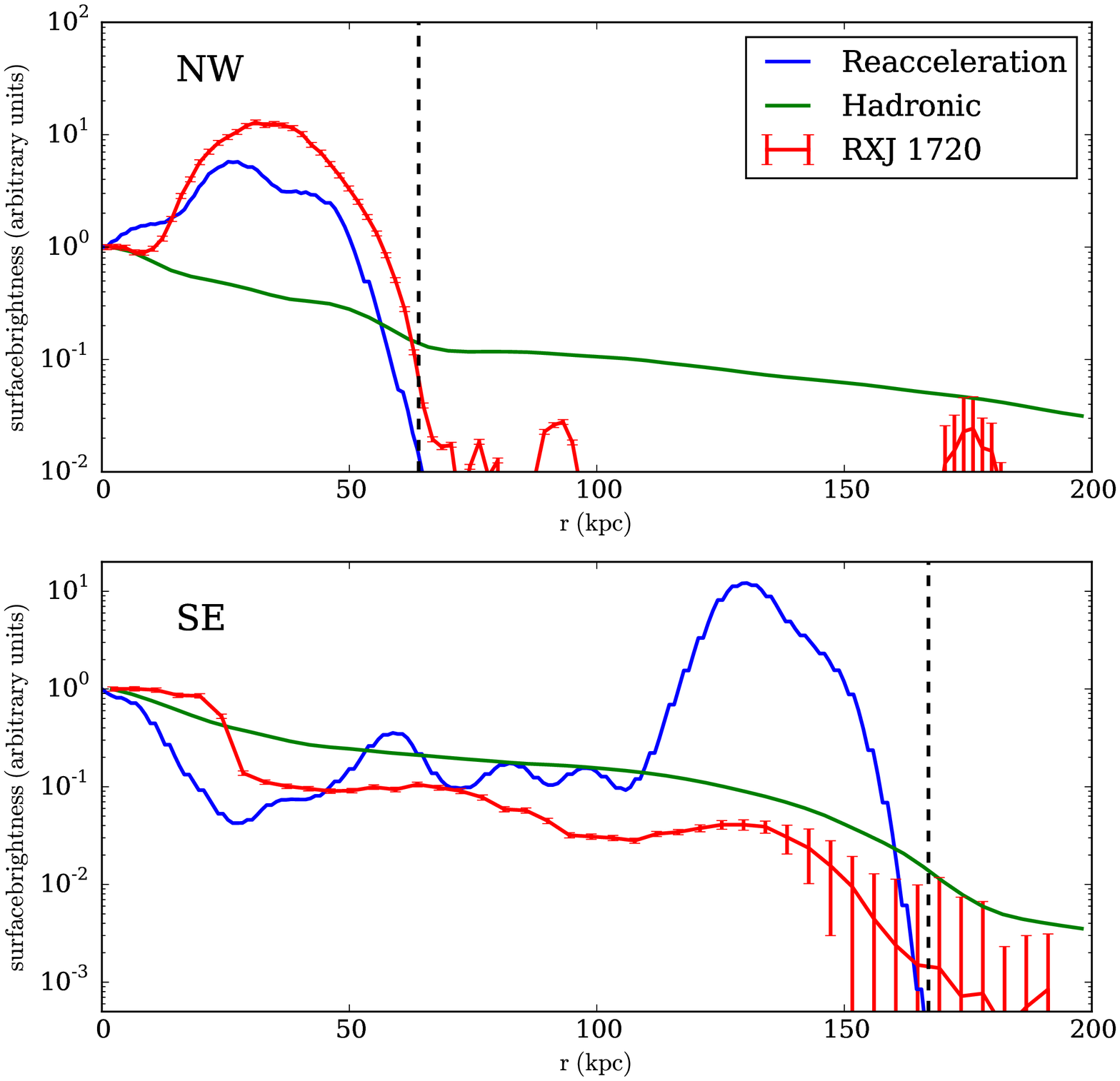}
\quad
\includegraphics[width=0.40\textwidth]{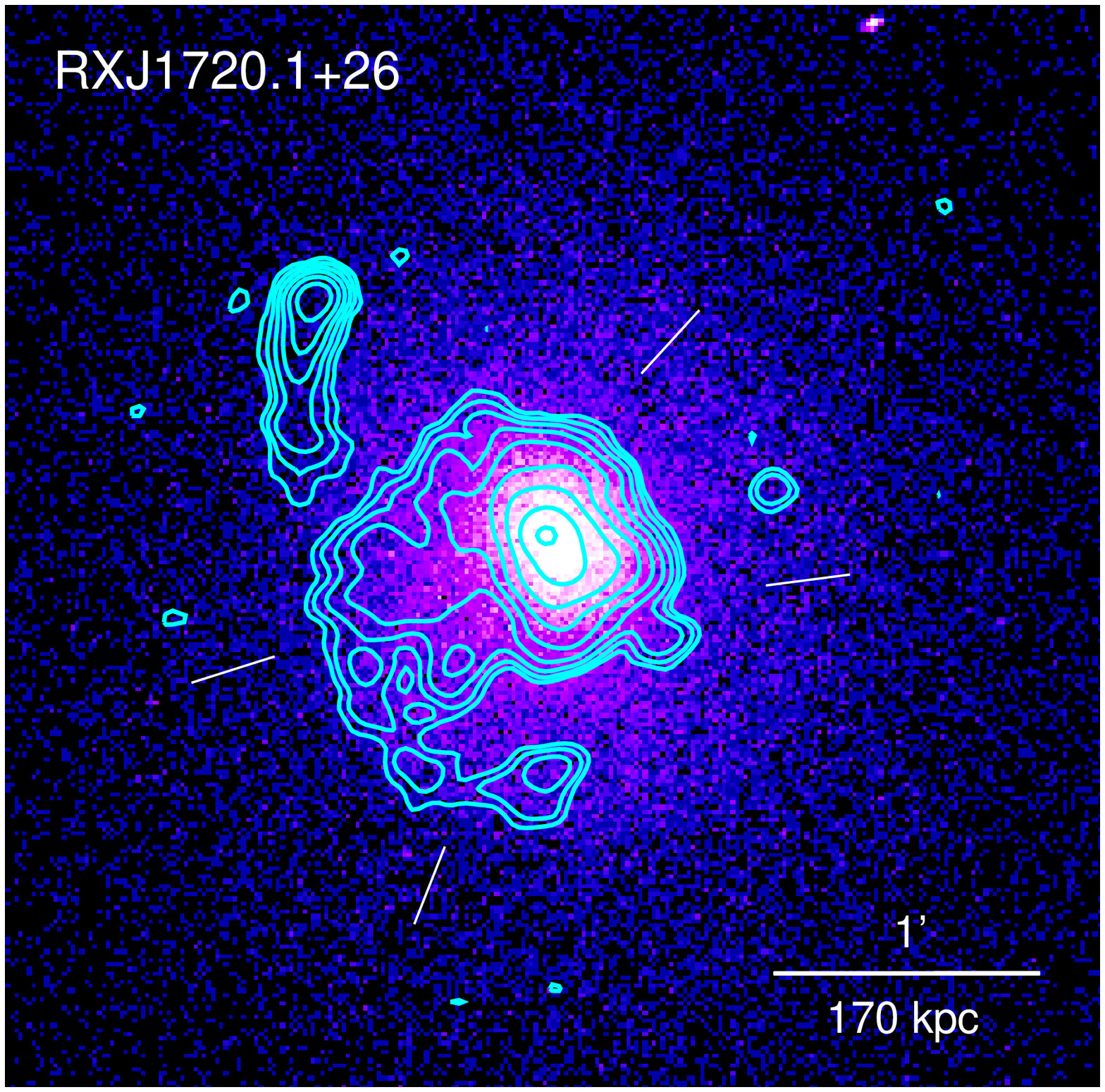}
\caption{Comparisons of radial profiles of radio emission between
  different models of minihalos and an observation. Left: Profiles of
  radio emission from the hadronic simulations from
  this work, the reacceleration simulations from Z13, and from
  observations of RXJ 1720.1+26 \citep[cf.][Figure 11]{gia14b}, along the northwest
  and southeast directions. The emission has been
  renormalized so that all profiles are unity at $r = 0$ for
  comparison purposes. The profiles from RXJ 1720.1+26 have been
  rescaled such that the radius of the cold front is the same as that
  in the simulations, and the error bars on these profiles are 1-$\sigma$. 
  The black dashed lines mark the positions of the cold fronts. Right: 
  X-ray counts image of RXJ 1720.1+26 with 617~MHz radio contours overlaid. 
  Radial white lines mark the two sectors (NW and SE) where the profiles have 
  been generated. The peak in the reacceleration profile at $r \sim$~130~kpc 
  is due to a patch of strong turbulence in this region.\label{fig:compare_profiles}}
\end{center}
\end{figure*}

Some observed minihalos \citep[e.g., RXJ 1720.1+26, MS 1455.0+2232][]{maz08,gia14b}
typically have well-defined edges or appear to be confined to the core
region, whereas others are more amorphous and/or extended \citep[e.g.,
A2029, Ophiuchus][]{gov09,mur10}. This potentially tells us something about the origin and /or transport of the relativistic particles. Unfortunately, in this work we have only simulated one cluster configuration, so we do not have any statistics for comparison, but we can at least provide some indications of what may be expected for future study. In our reacceleration simulation of Z13, the radio emission was confined completely within the cold fronts, whereas the radio emission in our hadronic model is more extensive, particularly in the northern direction. To illustrate this, Figure \ref{fig:compare_profiles} shows profiles of the radio emission
along the northwest and southeast profiles from the hadronic
simulation and the reacceleration simulation of Z13. Also plotted for
comparison are 617~MHz radial profiles of the radio emission of RXJ
1720.1+26 \citep{gia14b}, a galaxy cluster with sloshing cold fronts and a
minihalo qualitatively similar in shape to that from our simulations. The radii
of the profiles from RXJ 1720.1+26 have been rescaled such that the
radii of the X-ray cold fronts are the same as in the simulations. Both the
hadronic and reacceleration models result in drops in radio emission
at the cold front profiles, but the drops in the reacceleration model
are much steeper, and appear very similar to the drops in radio
emission in RXJ 1720.1+26. There is a strong increase in the
reacceleration profile at $r \sim$~130~kpc due to a patch of strong
turbulence in this region, but even without this feature the drop in
radio emission would be very pronounced. These results indicate that different physics potentially produces distinguishable predictions for the radial profiles of minihalos that can serve as a novel test to discriminate between different models for the origin of the emitting CRe.

\subsection{Spectral Index\label{sec:spec_index}}

Measurements of the spectral index provide crucial information on the acceleration and transport physics of the CR. In the previous section, we have shown that in the hadronic model that the vast majority of our tracer particles have essentially stationary electron spectra, and that consequently the synchrotron spectrum only depends on the input spectrum of CRp. This is different from reacceleration models, where turbulence plays a role in shaping the electron/synchrotron spectum and non-stationary conditions apply (by definition). Consequently the study of spectra is a powerful test to discriminate between the predictions of these two models, though in practice having multiple observations ranging over more than 1 decade in frequency will be necessary to start distinguishing between different models.

\begin{figure}
\begin{center}
\plotone{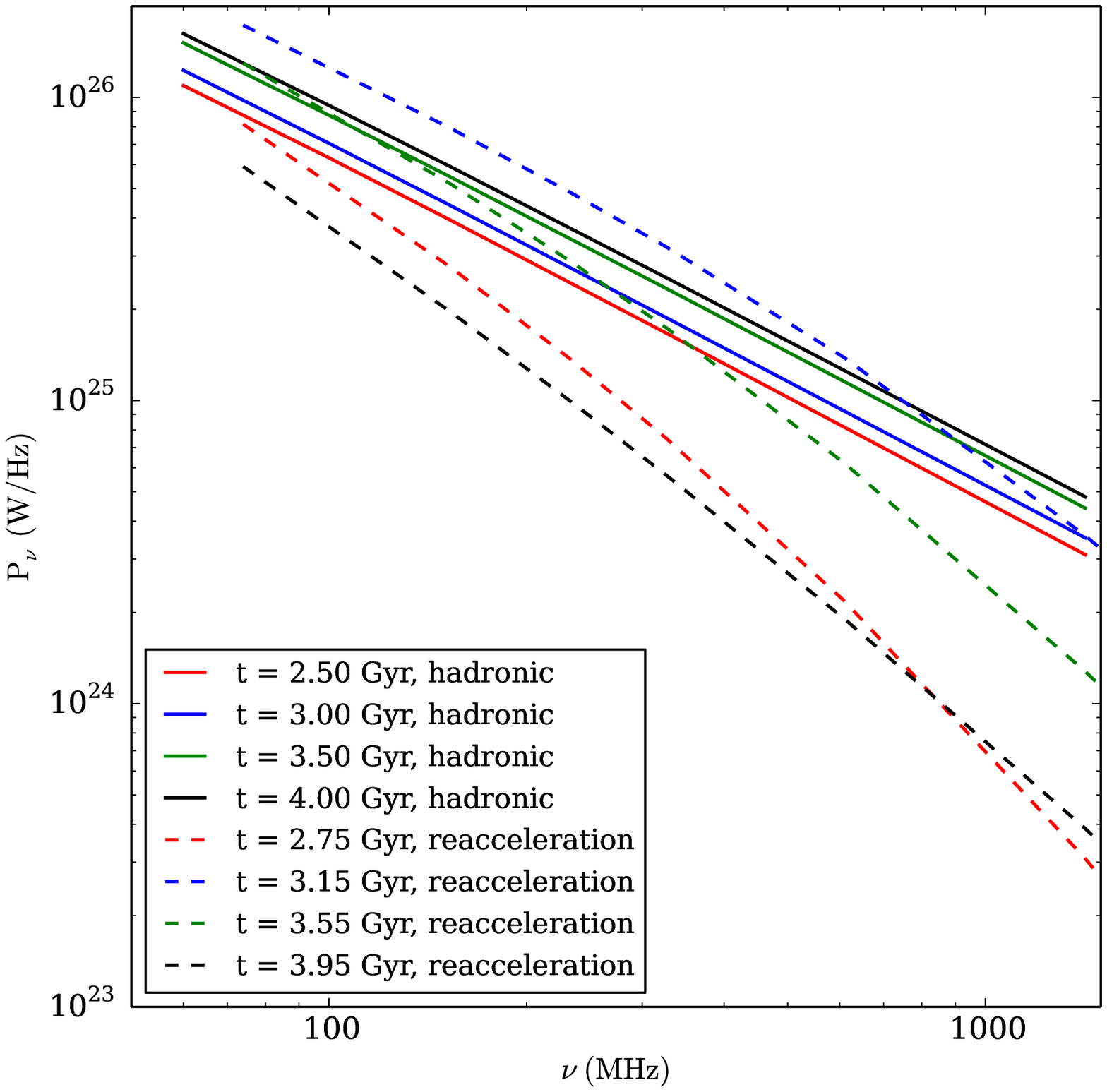}
\caption{The total radio spectrum within a radius of 300~kpc from the
  cluster center for the hadronic model from this work and the
  reacceleration model from Z13 at several epochs.\label{fig:compare_spectra}}
\end{center}
\end{figure}

Figure \ref{fig:compare_spectra} shows the total radio spectrum in W/Hz
within a radius of 300~kpc from the cluster potential minimum at
several different epochs. Spectra from Z13 at similar epochs for the
reacceleration model are also plotted for comparison. Our hadronic spectra are essentially power-law in shape over a large range of radio frequency, with the spectral index near the steady-state value $\alpha \approx 1.13$. In contrast, the reacceleration spectra gently steepen at high frequency, with different slopes depending on the epoch. 

\begin{figure*}
\begin{center}
\includegraphics[width=0.45\linewidth]{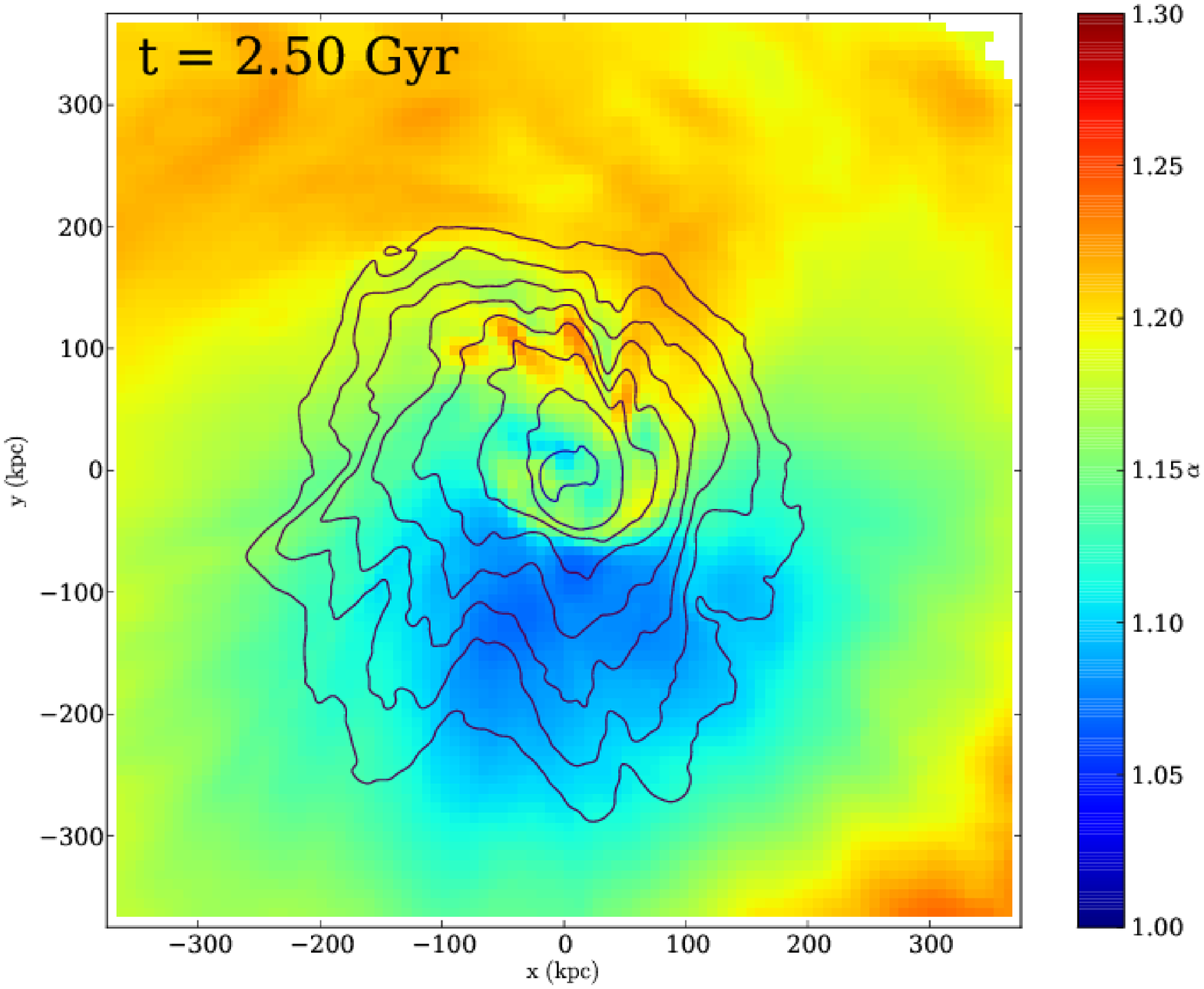}
\includegraphics[width=0.45\linewidth]{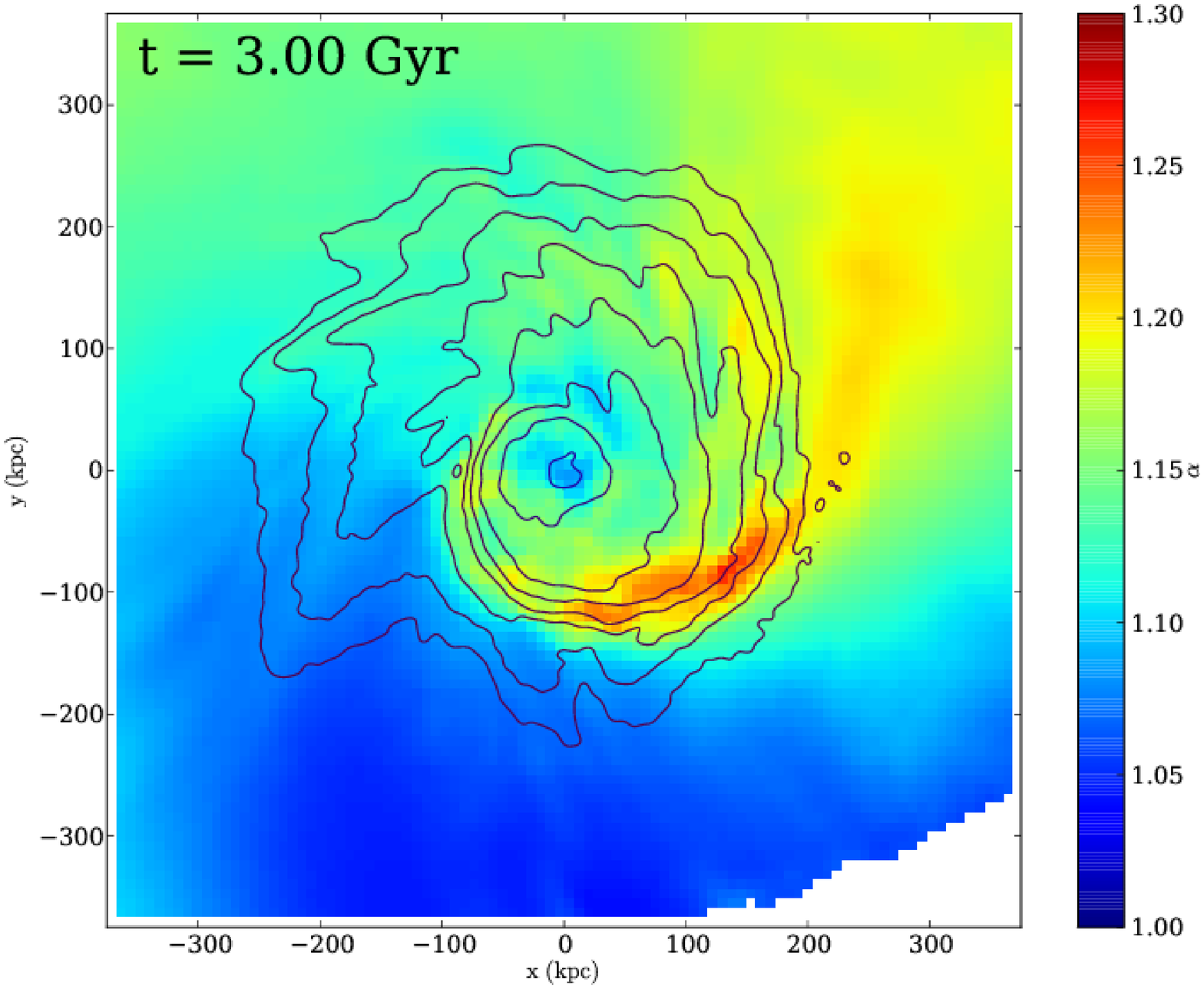}
\includegraphics[width=0.45\linewidth]{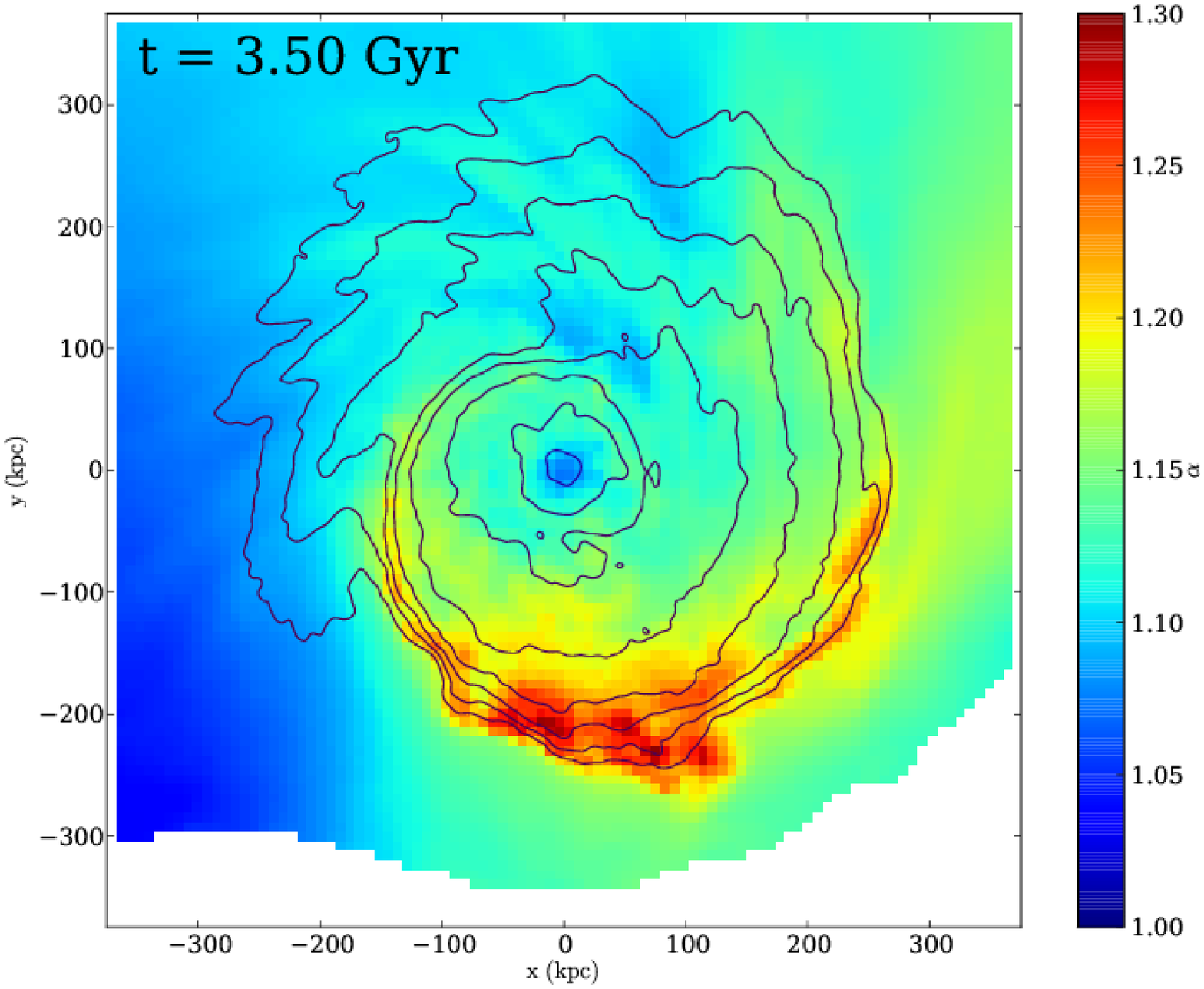}
\includegraphics[width=0.45\linewidth]{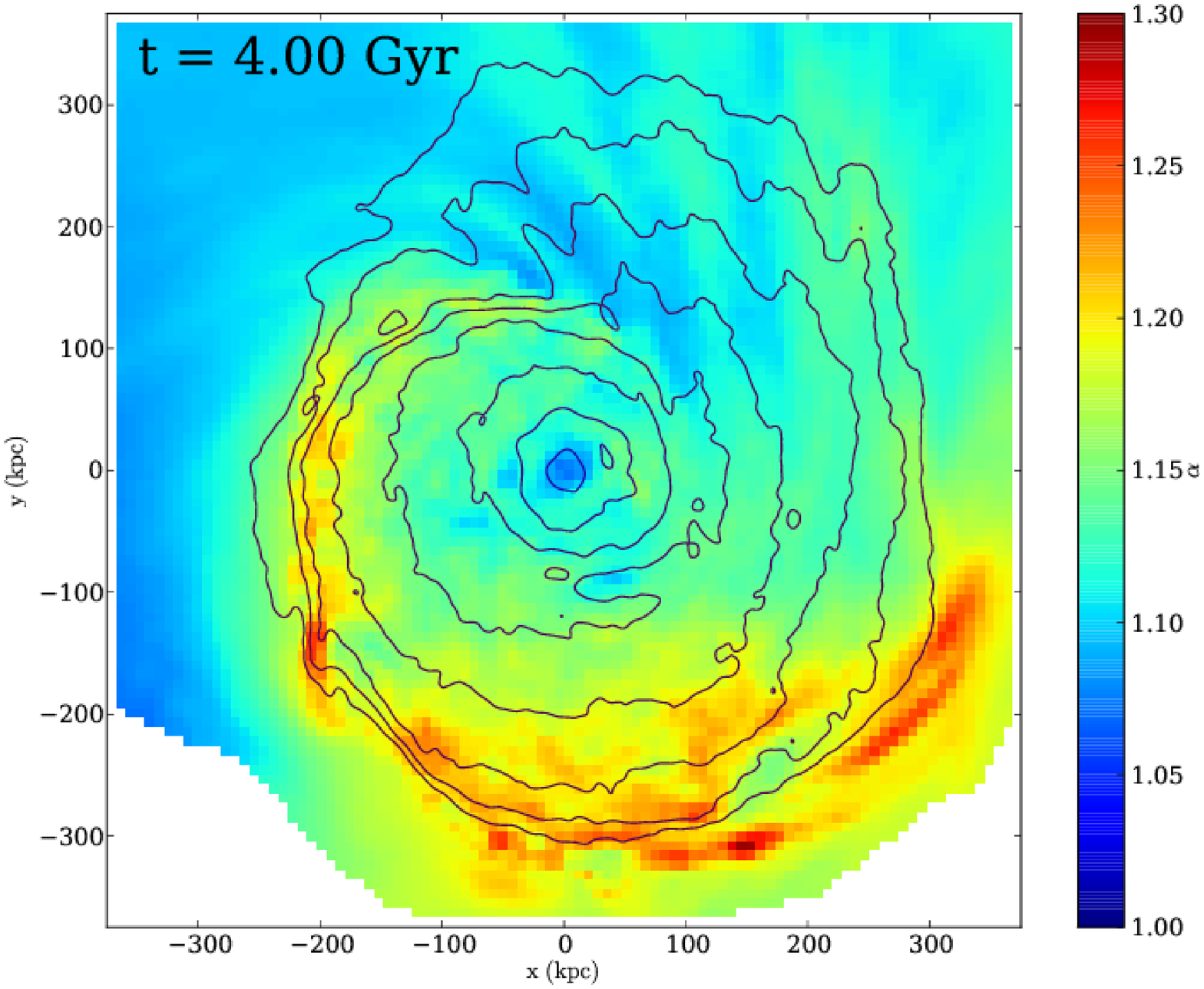}
\caption{Spectral index map at several epochs (between the frequencies of 327 and
  1420~MHz) with 327~MHz radio contours overlaid (contours are taken
  from Figure \ref{fig:327_maps}). Each panel is 750~kpc on a side.\label{fig:spec_index_maps}}
\end{center}
\end{figure*}

To create maps of the spectral index $\alpha$, we reblock our
intensity images by a factor of 8 and fit the spectrum in each
reblocked pixel within a frequency range to a power-law, assuming
10\% errors on the radio intensity in each band \citep[typical
  errors for the radio intensity in minihalo observations are within
  the $\sim$5-10\% range][]{gia14a}. Figure
\ref{fig:spec_index_maps} shows maps of the spectral index fitted
between 327 and 1420~MHz at several epochs of the simulation. We find a fairly uniform spectral index throughout the sloshing region, close to the steady-state expectation
of $\alpha \approx 1.13$, with the slopes in a narrow range of $\alpha
\approx 1.0-1.3$. Near the cluster center, the spectrum
becomes slightly flatter than the steady-state expectation ($\alpha
\approx 1$), consistent with the increasing importance of Coulomb
losses in the higher density region. Near the cold front edges,
where the magnetic field is increasing rapidly due to shear amplification, we
find that the spectral index increases to a maximum of $\alpha \approx
1.3$. This range of observed $\alpha$, roughly within $\Delta\alpha \sim 0.1-0.15$ is well within the observational uncertainties of the spectral index, e.g. Figure 6 of \citet{mur10}, and Figure 8 of \citet{gia14b}, indicating that when filtered by observational responses, this map might be difficult to distinguish from a spatially uniform distribution of spectral indices. 

Though we found that the spectral index for individual tracer particles in
these highly magnetized regions could be as steep as $\alpha \approx 2$ (see Section
\ref{sec:tracer_particles}), there are simply too few of these
steep-spectrum particles to affect the spectrum of the emission projected along the line of sight. These results are
inconsistent with the argument of \citet{kes10b} for minihalos, who
argued that rapid increases in magnetic field strength could explain
steep-spectrum radio emission. 

\begin{figure*}
\begin{center}
\includegraphics[width=0.45\linewidth]{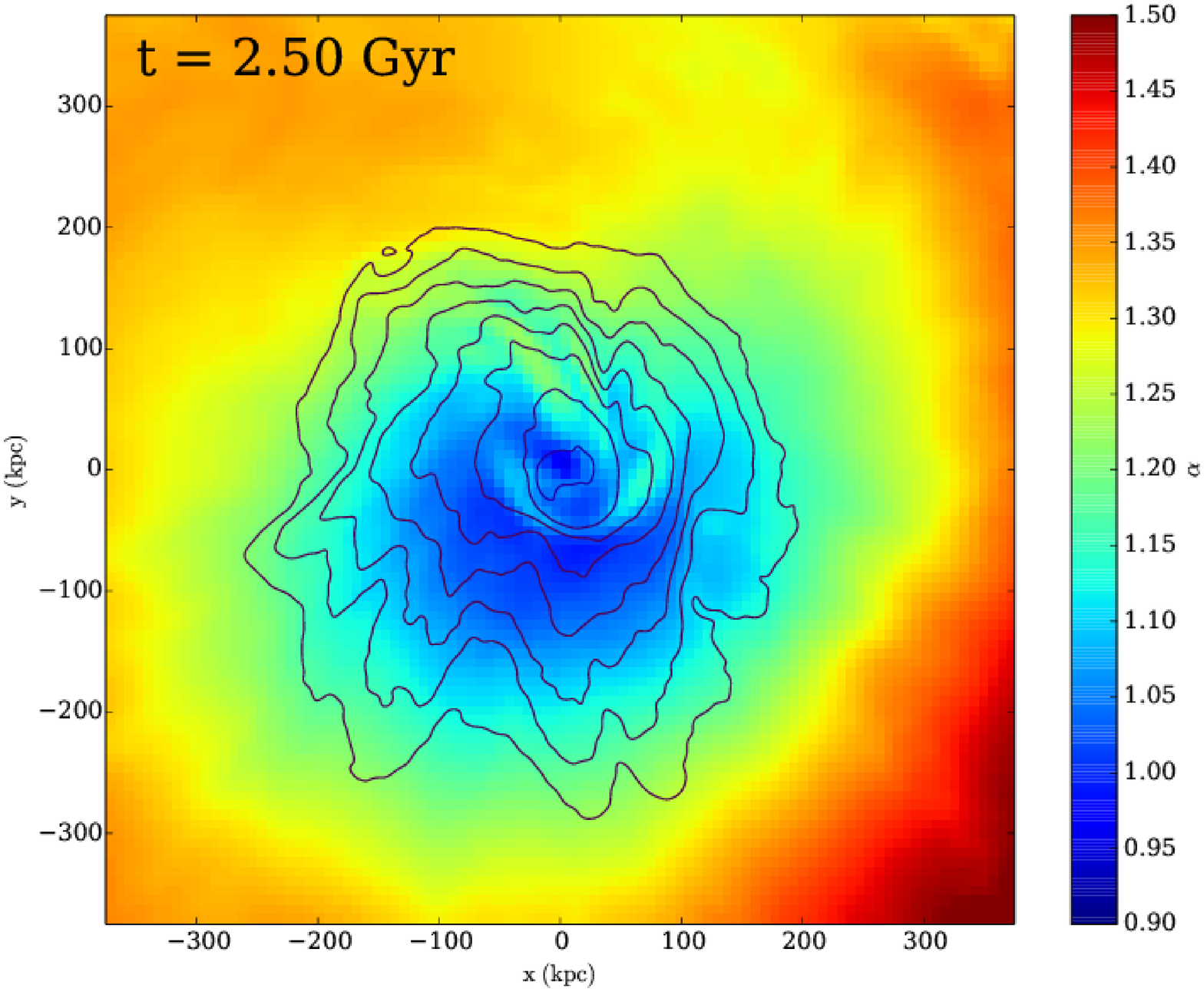}
\includegraphics[width=0.45\linewidth]{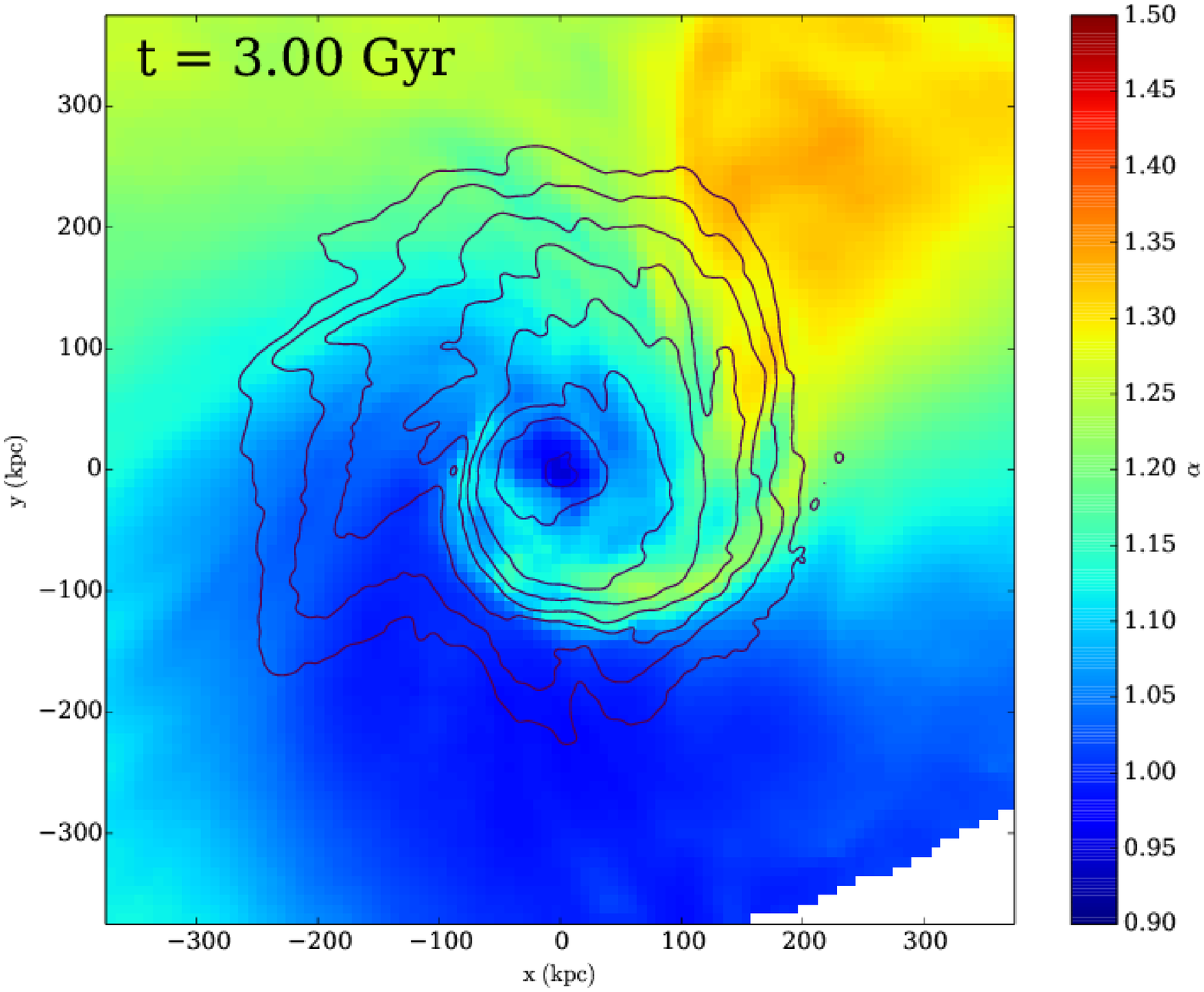}
\includegraphics[width=0.45\linewidth]{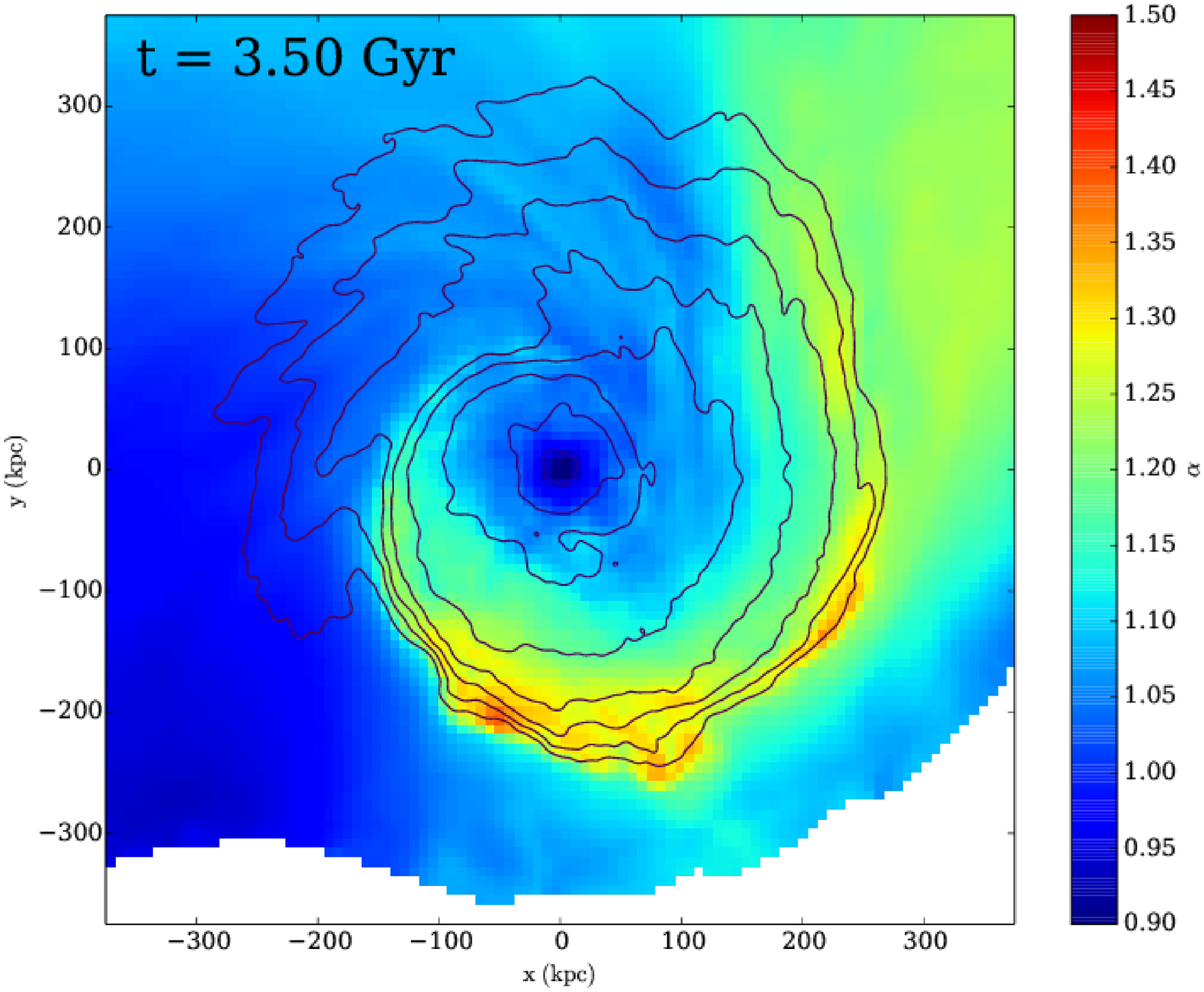}
\includegraphics[width=0.45\linewidth]{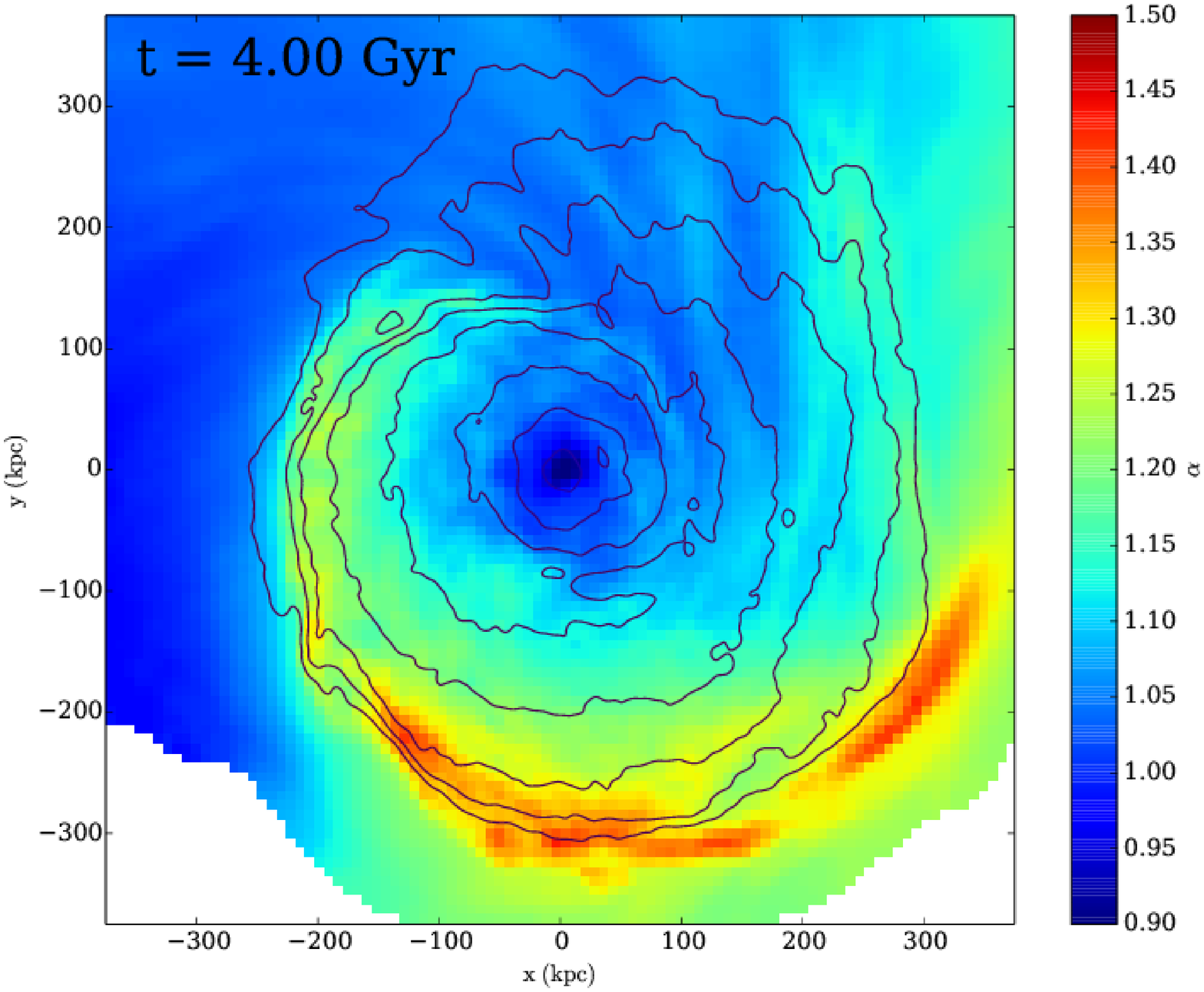}
\caption{Spectral index map at several epochs (between the frequencies of 60 and
  153~MHz) with 327~MHz radio contours overlaid (contours are taken
  from Figure \ref{fig:327_maps}). Each panel is 750~kpc on a side.\label{fig:spec_index_maps_low}}
\end{center}
\end{figure*}

LOFAR\footnote{http://www.lofar.org} and
LWA\footnote{http://www.phys.unm.edu/$\sim$lwa/index.html} will observe
radio sources at much lower frequencies, down to $\sim$40~MHz. At these frequencies, the
radiative loss timescales for the corresponding electrons are much
longer and the effects of a change in the magnetic field would be more
long-lasting. Figure \ref{fig:spec_index_maps_low} shows maps of the spectral
index fitted between 60 and 153~MHz at the same epochs of the
simulation. Here, there is again steepening of the radio spectrum at
the cold front surface, but this too is limited, up to $\alpha \approx 1.4$, only slightly
stronger than $\alpha$ measured in the higher-frequency band. At early epochs ($t \sim
2.5$~Gyr), we also see some marginal steepening of the radio spectrum just outside
the core region ($\sim 200$~kpc), where the magnetic field has been
recently strongly amplified due to the recent passage of the
subcluster and resulting compression of the plasma.  

\section{Discussion}

\subsection{The Radial Extent of Minihalos}\label{sec:radial_extent}

In some observed minihalos, the radio emission is strongly confined to the core
region, or the envelope of the cold fronts, if they are clearly observed.
In our simulations, we find that the radial profile of the minihalo emission
typically decreases at a more rapid rate approaching the cold fronts,
but this is not always the case (in particular the NW profile from
Figure \ref{fig:profiles}).

A drop in radio emission across a cold front in the secondary model
primarily results from two effects: the first is the density jump
across the cold front, which is typically a factor of $\sim$1.5-2 for
most cold fronts. The contribution of the density jump to the radio
drop will be modest, since for the case where the CRp energy density
is proportional to the thermal density, $j_\nu \propto n_{\rm
  th}P_{\rm th}$, and the thermal pressure is roughly continuous across
cold front surfaces (except in a very thin layer at the cold front
surfaces where the enhanced magnetic pressure due to shear
amplification may become significant). 

The main contribution to the drop in radio emission will arise from
the difference in the magnetic field across the front \citep{kes10a}. Below the
front, the field will be amplified to $B > B_{\rm CMB}$, and the radio emission is roughly
independent of the magnetic field strength. When the fronts reach a
large radius, above the front the unamplified field will likely be
weak due to the decline of magnetic field strength with radius, with $B < B_{\rm
  CMB}$, and the radio emission will drop off rapidly as the square of
the field. Figure \ref{fig:scaledB} shows the mass-weighted projected
magnetic field at the epoch $t = 3.5$~Gyr scaled by $B_{\rm CMB}$. Throughout the volume of the sloshing region, the magnetic field has been amplified, and is stronger than $B_{\rm CMB}$. For the most part, just outside of the cold
fronts the field drops to $B < B_{\rm CMB}$. However, the region to
the northwest is filled with strong magnetic field outside of the cold
fronts, due to strong velocity shears in that region. Along
projected radii in that direction, there will be no drop in emission
as one crosses the cold front surface. Clearly, in the hadronic model
the radial extent of the minihalo emission depends critically on the magnetic field structure. 

This, of course, depends to a certain extent on our initial choice for
the magnetic field strength. In some clusters, the magnetic
field strength may be even higher than in our setup (given by $\beta \sim 100$, see Section \ref{sec:MHD}) due to turbulent-driven amplification caused by AGN, minor mergers, and accretion. More simulations with different magnetic field configurations would be needed in order to determine the conditions under which minihalos would have steep drops in emission at cold front surfaces in the hadronic model. The significance of the drop in emission due to a sharp decrease in the magnetic field strength across the front is also affected by the strong redshift dependence of the CMB energy density, with $B_{\rm CMB} \propto (1+z)^2$.

\subsection{Total Spectrum and Spectral Steepening Along the Cold Fronts}

According to \citet{kes10b}, rapid magnetic field amplification can
cause significant departures from steady-state conditions, which
produces synchrotron spectral steepening under the hypothesis of a
secondary origin of the emitting particles. If strong, these effects
may have important consequences for interpreting the origin of minihalos and giant
halos. In our simulations we do not see significant effects of this
nature. We find that the total spectrum of the minihalo is fairly consistent with a simple
power-law in agreement with the expectation based on steady-state
conditions. Though we do see some steepening of the synchrotron
spectrum near the cold front surfaces that we can confirm is indeed
due to magnetic field amplification in these regions, we find that the
effect is barely observable, even at very low observing frequencies where it should be
maximized. Section 4.3.3 of \citet{kes10b} determines the change in
spectral index $\alpha$ for a sudden increase in the cooling rate over
the injection rate, parameterized by
\begin{equation}
\R = \frac{\psi_f/\Q_f}{\psi_i/\Q_i} = \frac{B_f^2+B_{\rm CMB,f}^2}{B_i^2+B_{\rm CMB,i}^2}\frac{n_{\rm
    th,i}C_{\rm p,i}}{n_{\rm th,f}C_{\rm p,f}}.
\end{equation}
From Figure 16 of \citet{kes10b}, we may determine the approximate
increase in the magnetic field strength associated with a given
spectral steepening, which will serve as an independent check on our model for the steepening of the CRe spectra. It is instructive to examine the necessary
increase in the cooling rate to produce the steepest spectra of any of
the particles in the simulation ($\alpha \sim 2$) and the steepest
spectra we measure in our spectral index maps ($\alpha \sim 1.3$). 
For our injection index of $\alpha_p \approx 2$, a sudden increase of
$\R \sim 15$ is required to steepen the synchrotron spectrum to
$\alpha \sim 2$, whereas for steepening to $\alpha \sim 1.3$ a more
modest increase of $\R \sim 3$ is required. 

What qualifies as a ``sudden'' increase? Formally, since in
steady-state conditions the characteristic timescale is \citep{kes10a}:
\begin{equation}
t_{\rm cool} \simeq 0.13\left[\frac{4\left(\frac{B\sqrt{3}}{B_{\rm
        CMB,0}}\right)^{-3/2}}{1+\left(\frac{B}{B_{\rm CMB,0}}\right)^{-2}}\right]\nu_{1.4}^{-\frac{1}{2}}(1+z)^{-\frac{7}{2}}~{\rm Gyr},
\end{equation}
any sudden increase in the cooling rate must occur on a
timescale shorter than this in order to significantly affect the
spectrum. For $B \simlt B_{\rm CMB}$, $t_{\rm cool} \sim 0.1-0.2$~Gyr
for electrons emitting at frequencies between 327 and 1400~MHz. 

Motivated by this, we examine the trajectories of a few tracer
particles with extreme ($\alpha \sim 2$) spectral indices, to see if
these conditions prevail. Figure \ref{fig:cooling_evolution} shows the evolution in
$\R$ for three tracer particles with spectral indices $\alpha \sim
2$. $\R$ for each particle is normalized to unity at time $t_i$, and all times are given
with respect to the epoch $t_f$ at which the spectral index is
measured, marked by the black dashed line. Each of these particles indeed experiences a sudden increase in $\R$, followed by a rapid decrease. The dot-dashed lines in the
figure indicate the necessary $\R$ to steepen the spectrum to the
given spectral index from a steady-state, estimated from Figure 16 in
\citet{kes10b}. All of the curves show an increase in $\R$ that is at
least this much, and this occurs over roughly a cooling time ($t \sim 0.13$~Gyr), with the most significant increase occurring over a small fraction of that time. We therefore find that the evolution of the CRe energies for the steepest-spectrum tracer particles in our simulation is consistent with the results of \citep{kes10b}.

\begin{figure}
\begin{center}
\plotone{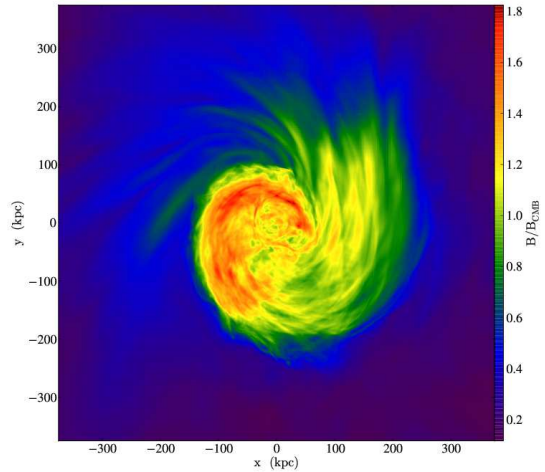}
\caption{Mass-weighted projection of the magnetic field strength
  scaled by $B_{\rm CMB}$. The panel is 750~kpc on a side.\label{fig:scaledB}}
\end{center}
\end{figure}

\begin{figure}[!t]
\begin{center}
\plotone{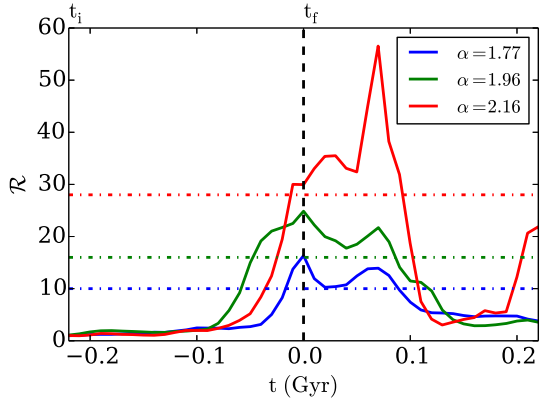}
\caption{Evolution of the ratio of the cooling rate to the injection
  rate $\R$ for three different tracer particles with extreme spectral
  indices. Times are rescaled to $t_f$ (for each particle), the epoch at which the spectral
  index is measured, marked by the black dashed line. The ratio $\R$ is rescaled so that its value is
  unity at $t_i$.  The colored dot-dashed lines indicate the value of
  $\R$ for the corresponding particle required to steepen its spectrum
  to its spectral index value at $t_f$, estimated from Figure 16 of
  \citet{kes10b}. \label{fig:cooling_evolution}}
\end{center}
\end{figure}

The fact that we find so few tracer particles with very steep spectra
indicates that these conditions (required to produce such steepening) do not prevail
throughout most of the cluster core. Though there is significant
magnetic field amplification (see Figure \ref{fig:scaledB}), it does not usually occur fast enough to steepen the spectrum significantly. 

\subsection{Limitations of this Work}\label{sec:limitations}

In this work, we have adopted a simple model for hadronically originating CRe, by relaxing the typical steady-state assumption and allowing for the time evolution of the CRe spectrum due to changes in the local gas and CRp density and the magnetic field strength, as well as the evolution of the CMB energy density with redshift. Relaxing the steady-state assumption is meaningful in our simulation, due to our high spatial and time resolution which are able to resolve adequately the amplification of the magnetic field. 

As a significant simplification, we also assumed that at any given epoch the energy density of CRp was proportional to the energy density of the thermal gas, and that the input spectrum of the CRp is a constant power-law spectrum (excepting the cutoff at the low-energy threshold). This approach is therefore limited, as it relies on a simplified treatment of the spatial distribution and energetics of the CRp.

First, we are not able to model in a self-consistent fashion the transport and diffusion of CRp and its connection with the gas dynamics in the sloshing
region. Turbulence might transport CRp on a scale that could be of the
order of the cluster core, potentially inducing a spatial distribution
of CRp that is broader than that of the thermal plasma. Momentum-diffusion of CRs that is mediated by the scattering of particles with MHD waves in the plasma is a different process that is more difficult to model due to the uncertainties in the physics of
(small-scale) turbulence in these environments. In addition, \citet{ens11} and \citet{wie13} proposed that super-Alfv\'enic CR streaming is possible in the ICM. In
particular, \citet{wie13} calculated the suppression of the streaming
instability in a turbulent flow showing that under these conditions
the self-generated waves do not limit the particle drift velocity to
the Alfv\'en speed. However, this does not automatically imply that
streaming is efficient, because the background turbulence (necessary
to suppress the instability) still provides a source of scattering
that may make the transport of CRs diffusive and potentially
inefficient \citep[see][for a discussion]{bru14}. The combination of these effects may significantly affect the expected properties of minihalos with respect to our simplified treatment. Turbulent transport and streaming would be the most important effects, and they typically produce a broader spatial distribution of CRp with respect to our simplified model. However, if these effects are important, they would produce synchrotron distribution even broader than those obtained by our model, making the difference between reacceleration and hadronic models even more relevant. In particular, the radial drops in hadronic models could become even smoother, increasing the tension with observations. Another possibility that we cannot rule out, which would produce the opposite effect, is that the CRp which are ultimately responsible for the radio emission are generated in the cluster center (e.g., from AGN activity), and then diffuse outward, but are confined to the region bounded by the cold fronts due to the fact that the field direction is largely tangential at the front surfaces (ZML11) and thus would prevent diffusion across them. To investigate these and other effects, future studies of sloshing and minihalos based on simulation methods that self-consistently follow the transport of CRp will be needed.

Second, given the simplified assumptions in our model with respect to the CRp, the spatial and spectral distribution of the radio emission that is produced in our model is determined almost entirely by the spatial distribution and time evolution of the magnetic field. It is therefore crucial that we adequately resolve this evolution and are not missing any significant sources of further amplification. 

The magnetic field in our cluster core will be amplified by shearing motions localized at the cold fronts, as well as turbulence generated by sloshing. \citet{kes10} identified shear flows at cold front surfaces as the most relevant mechanism for the amplification of the magnetic field, which was confirmed in ideal MHD simulations by ZML11. However, turbulence will also play a role. In our inviscid simulation, turbulent dissipation is entirely numerical, and determined by the spatial resolution. Therefore, though we resolve gas motions down to scales of $\sim$kpc, the scale at which dissipation of the turbulent cascade begins to set in is slightly larger, around $\sim$10~kpc (see Z13 and references therein). 

In any case, any further field amplification provided by turbulent motions on scales that are not resolved by our simulations is likely to be a small fraction of what we are able to accomplish with our high-resolution simulation. The turbulence is injected via the sloshing motions on scales of $\sim$100~kpc, and cascades down to the previously mentioned scales where it is dissipated by our finite resolution. The energy flux through the turbulent cascade, $\epsilon_t \propto v_\ell^3/\ell$ (assuming a Kolmogorov spectrum) is constant with respect to the scale $\ell$. Given this fact, the kinetic energy available for conversion into magnetic energy at these unresolved scales will be small compared with the energy available at the scales we do resolve. Therefore, we do not expect that there is any significant room for further amplification by turbulence by resolving these smaller scales. Even an increase in the magnetic field energy by a few tens of percent would not significantly change our conclusions. Additionally, our previous work (Z13) showed that the kinetic energy in turbulence generated by sloshing on scales less than $\sim$30~kpc is estimated to be at most a few percent of the thermal energy, and even if all of this energy were to be transferred to the magnetic field it would still not be as large as the contribution from the larger-scale sloshing motions (see Figure 24, Z13). In the Appendix, we use a resolution study to demonstrate that we have converged with respect to the overall properties of the magnetic field within the core region.

However, it is the case that our specific initial value for the overall ratio of the thermal to the magnetic pressure, $\beta$, is a free parameter in our simulations. ZML11 investigated a range of initial values for $\beta$ (over an order of magnitude from $\sim$100-6400), and demonstrated that the average magnetic field strength within the core region was similar at late times, regardless of the initial field strength. We may predict that for initial $\beta$ higher than our value of 100 (a weaker initial field), that the contrast between the field strength inside and outside the cold front would be larger, and there would be a sharper decrease in the radio emission at the cold front surface. On the other hand, for a lower initial $\beta$ (a stronger initial field), the contrast between the magnetic field strength inside and outside the front would be smaller, and the drop in radio emission would be less significant.

Third, we assumed a population of CRp with a single power-law
spectrum. This implies an injection spectrum of CRe that is power-law
in form, and that any steepening of the radio spectrum at higher
frequencies comes from strong departures from the steady-state
condition. Instead, steepening of the CRe spectrum may reflect
possible steepening of the CRp spectrum. Although the momentum distribution of CRs is expected to be a power law in a very broad range of astrophysical situations, this is certainly a simplification. In our conditions, momentum-dependent spatial transport of CRs could change the spectrum of CRp. For example, \citet{wie13} discussed the
possibility of a steepening of the CRp spectral distribution resulting
from an efficient momentum-dependent diffusion. This is also not
included in our simulations. However, in non steady-state conditions, this process would eventually result in a spectral flattening of the synchrotron emission with distance, because higher energy CRp would not be diffusively trapped and propagate faster faster (this effect is unavoidable in the case that we measure variations of the radio spectrum with distance). The observational consequence would be that minihalos
would be typically broader if observed at higher frequencies. This
differs from expectations based on other models, allowing to
discriminate between different origin scenarios for the CRe. For example, reacceleration
models eventually predict a spectral steepening with radius or more
complex spectral variations \citep[e.g.][and references
therein]{bru14}. Another limitation is that we do not include the
effect of turbulent reacceleration on the spectrum of both CRp and CRe, a process
that is believed to be important, for example, in the case of giant radio halos.

\section{Summary}

In this work, we have employed a MHD simulation of gas sloshing in a
galaxy cluster core to model the formation of a radio minihalo from
hadronically-originating cosmic-ray electrons. We used passive tracer particles to follow the evolution of CRe spectra under the assumption that these electrons are continuously injected by hadronic processes and they undergo radiative (synchrotron and inverse-Compton) and Coulomb losses via interaction with the thermal gas and magnetic fields. Our model represents a simple extension beyond that of a steady-state approximation between injection and loss processes in that fast changes in the magnetic field strength are allowed to change the shape of the electron spectrum.

The radio emission that is produced in our simulation is similar to that of observed minihalos in terms of being spatially diffuse and also being of similar luminosity. It also follows the overall spiral pattern of the cold fronts, expected due to the amplification of the magnetic field within these structures. In this respect, the model presented here and the results of our earlier work (Z13) using a reacceleration model give similar results.

Going beyond these considerations, the spatial extent and spectral steepening of radio minihalos provide avenues of distinguishing between models with different origin scenarios for the cosmic-ray electrons. Therefore, in this work our aim was to determine the effects of a strong, fast amplification of the magnetic field due to sloshing motions on these two properties of the radio emission. This model is attractive because it does not require any further assumptions about the input spectrum and spatial distribution of the cosmic-ray protons.

Under these assumptions, we find that the radio emission associated with the minihalo is not completely confined within the core region or the area bounded by the cold fronts. Along some radial directions that cross cold front surfaces, the radio emission drops steeply at these surfaces, due to the difference in the magnetic field strength across the interface. However, along other directions, the magnetic field strength changes less sharply at the fronts and as a result the radio emission decreases much more smoothly with radius, in concert with the mild decline in X-ray emission. 

We also find that some steepening of the synchrotron spectrum does occur in some regions due to rapid magnetic field amplification caused by sloshing. For some test particles, this effect is significant, with radio spectra as steep as $\alpha \approx
2$ (compared to the steady-state, and prevalent, value of $\alpha =
1.13$). However, the emission from these particles makes up a very
small fraction of the total emission from the CRe. As a result, we
find that the total spectrum of the minihalo is essentially a power law, consistent with that expected from assuming steady-state conditions. The effects of rapid magnetic field amplification are more visible when we consider synchrotron spectral index maps. In these, we see some steepening in the region of the cold fronts where magnetic
field amplification is stronger. However these effects induce only
marginal variations of $\Delta{\alpha} \sim$ 0.1-0.15 between 327-1420~MHz and
$\Delta{\alpha} \sim$ 0.2-0.25 between 60-153~MHz as one goes outward along the
cold fronts, where magnetic field amplification is strongest. This
differs from our previous results based on the same underlying numerical
simulation under the hypothesis that the emitting electrons are
reacccelerated by turbulence, where the spectral index is
predicted to steepen more than this at high frequencies. 

Based on our numerical studies from this work and Z13, we believe that the issues highlighted above regarding the spatial extent and the spectral slope of the minihalo emission are important to consider when discriminating between different origin models of the emitting CRe. For example, our expectations based on our hadronic model are in tension with observations of a few of the best-observed minihalos, which show the radio emission confined to roughly the size of the core region in all directions \citep{maz08,gia11,gia14a}. 
Similarly, spectral studies of radio minihalos \citep[such as
in][]{mur10,gia14b} indicate that significant spectral steepening is likely to be common in these sources. Though the few observations currently available do not permit us to make firm conclusions, addressing these differences between the two models should be a focus of further studies.

However, the parameter space explored in this work was necessarily limited. The most obvious extension of this work would need to incorporate the dynamics of the CRp self-consistently to determine to what extent it plays a role in shaping the radio emission from minihalos. Secondly, though our models resolve the structure of the magnetic field adequately, different initial conditions for the magnetic field strength and spatial configuration could be explored. Finally, a fuller picture of the cause of such emission may arise from the combination of the effects of hadronic injection of CRe and their subsequent turbulent reacceleration, a merger of the two currently competing
models, which has already been investigated for giant radio halos from a theoretical perspective \citep{bru05,bru11}.

\acknowledgments
JAZ thanks Uri Keshet for useful discussions. Analysis of the simulation data was carried out using the AMR analysis and visualization toolset yt \citep{tur11}, which is available for download at \url{http://yt-project.org}. JAZ is supported under the NASA Postdoctoral Program. GB acknowledges partial support by grant PRIN-INAF-2009. The software used in this work was in part developed by the DOE-supported ASC / Alliances Center for Astrophysical Thermonuclear Flashes at the University of Chicago.

\appendix

\section{Resolution Test: Convergence of the Magnetic Field\label{sec:bfield_convergence}}

\begin{figure*}
\begin{center}
\includegraphics[width=0.49\linewidth]{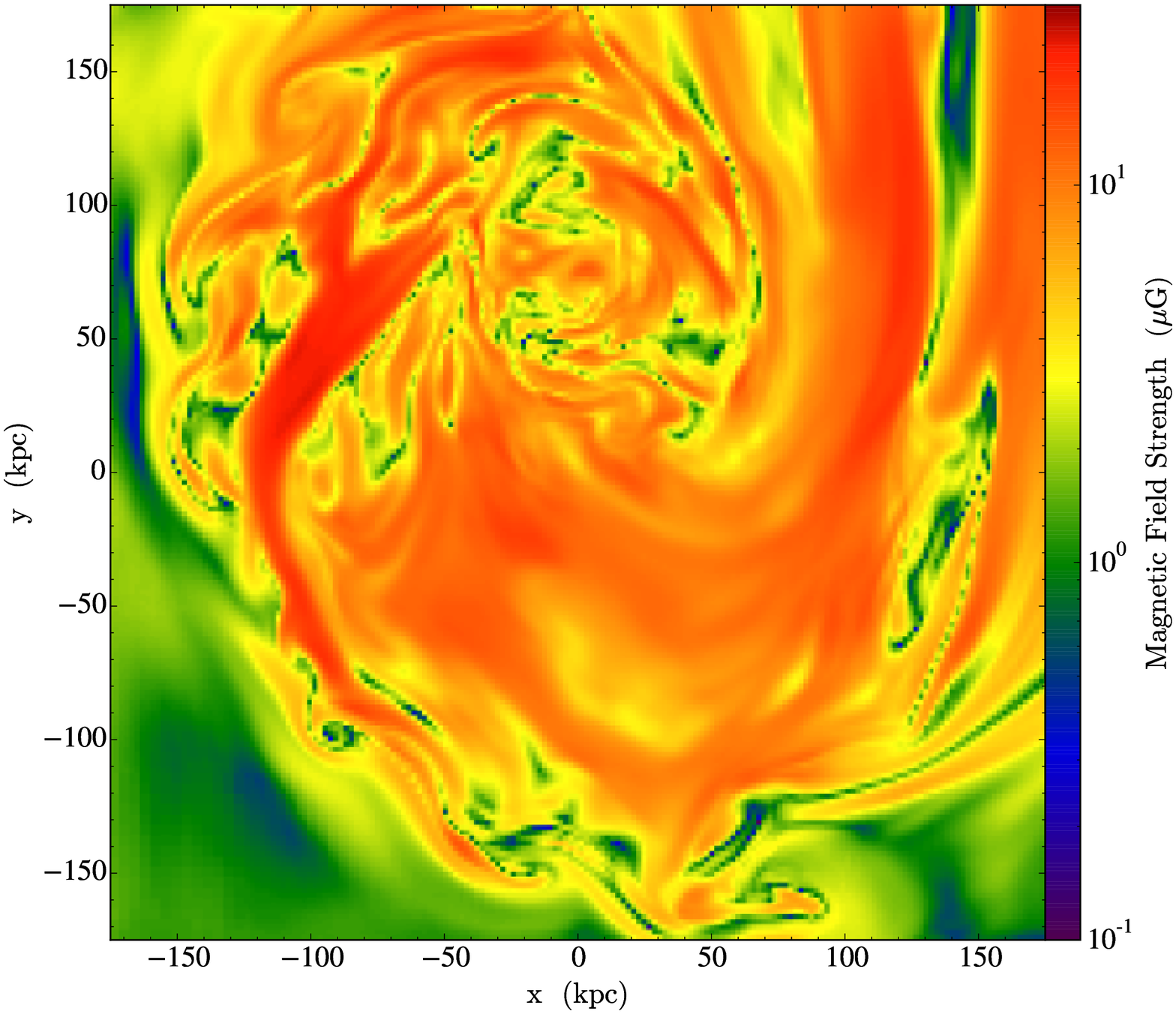}
\includegraphics[width=0.49\linewidth]{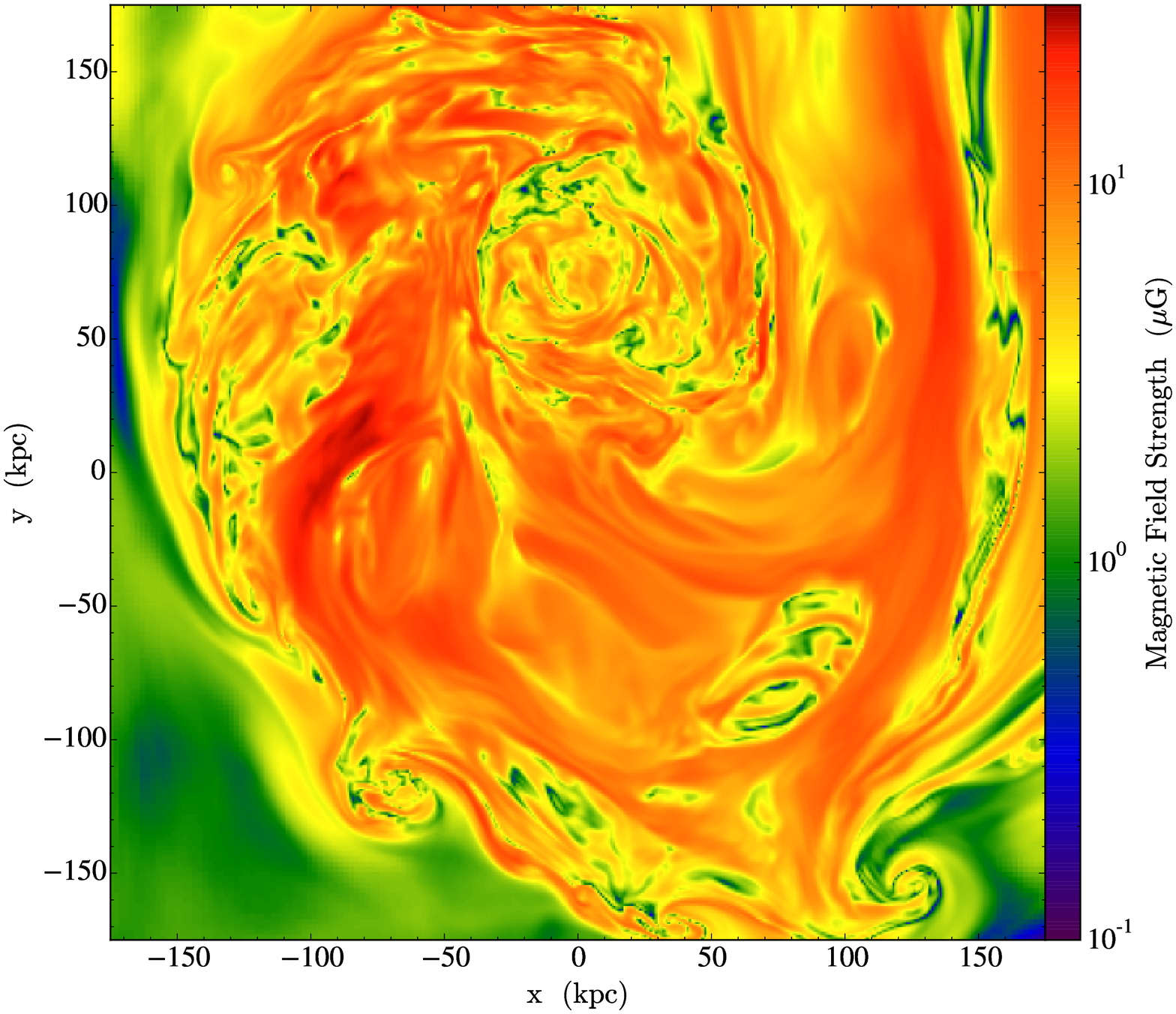}
\caption{Slices of the magnetic field strength through the domain center of the simulation from this work, at two different resolutions. Left: $\Delta{x} = 2$~kpc. Right: $\Delta{x} = 1$~kpc. Each panel is 350~kpc on a side.\label{fig:AM06_res_test}}
\end{center}
\end{figure*}

The finite resolution of our simulations limits the amplification of the magnetic field strength that can occur, either due to turbulence or bulk motions like sloshing (which is the predominant source of such amplification in our simulation). Higher resolution simulations will result in stronger magnetic fields in regions where vorticity or shear is significant. To examine the effect of resolution on our simulations, we compare our default simulation with a finest cell width of $\Delta{x} = 1$~kpc to an otherwise identical simulation that has been evolved with a finest cell width of $\Delta{x} = 2$~kpc. Figure \ref{fig:AM06_res_test} shows slices of the magnetic field strength through the center of the domain for both of these simulations at the epoch $t$ = 3.5~Gyr. The simulation with higher resolution exhibits more small-scale structure in the magnetic field, along with localized regions where the magnetic field is stronger than the corresponding regions in its lower-resolution counterpart. However, the average field over larger scales is the same in both simulations.

\begin{figure*}
\begin{center}
\includegraphics[width=0.49\linewidth]{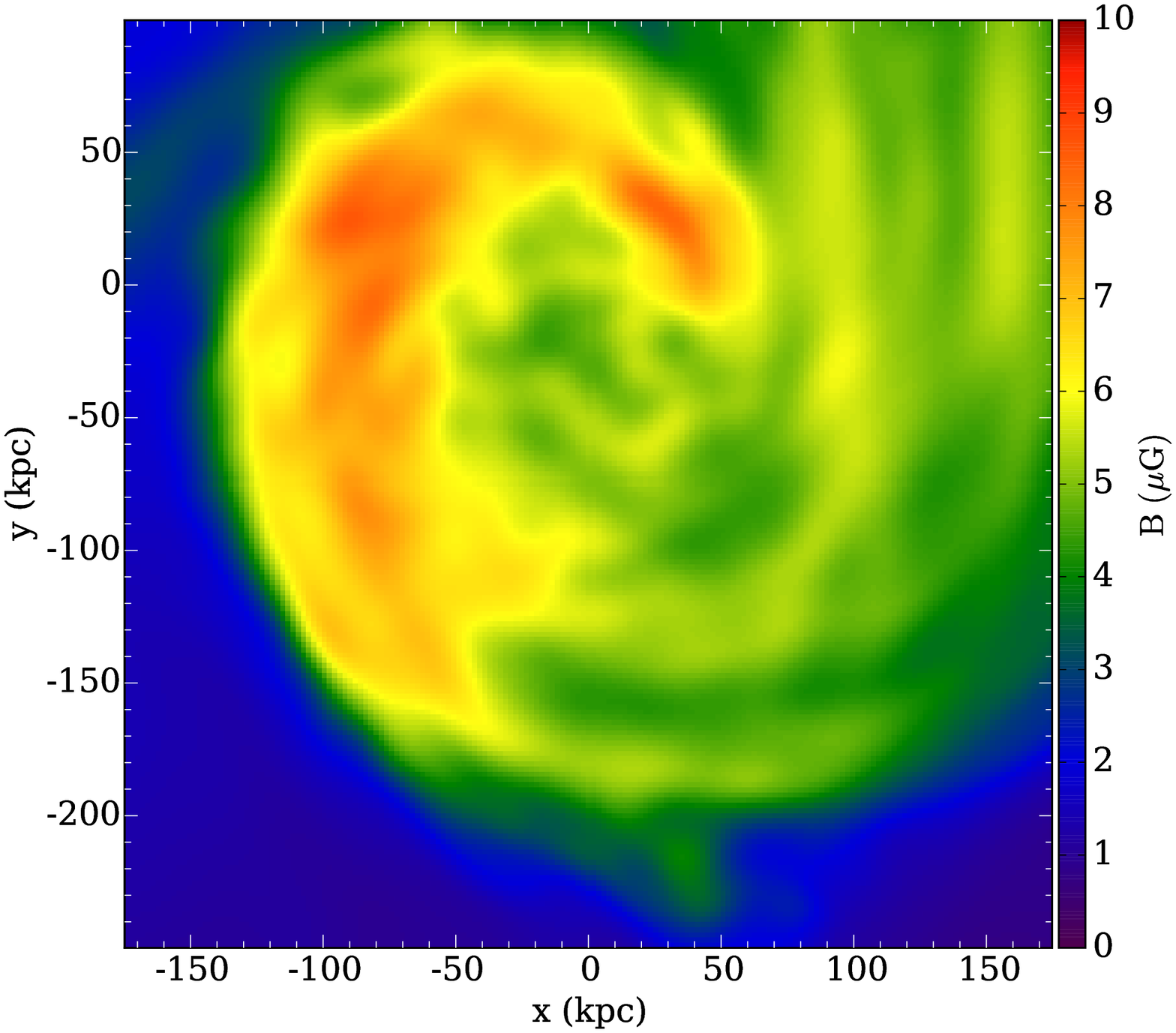}
\includegraphics[width=0.49\linewidth]{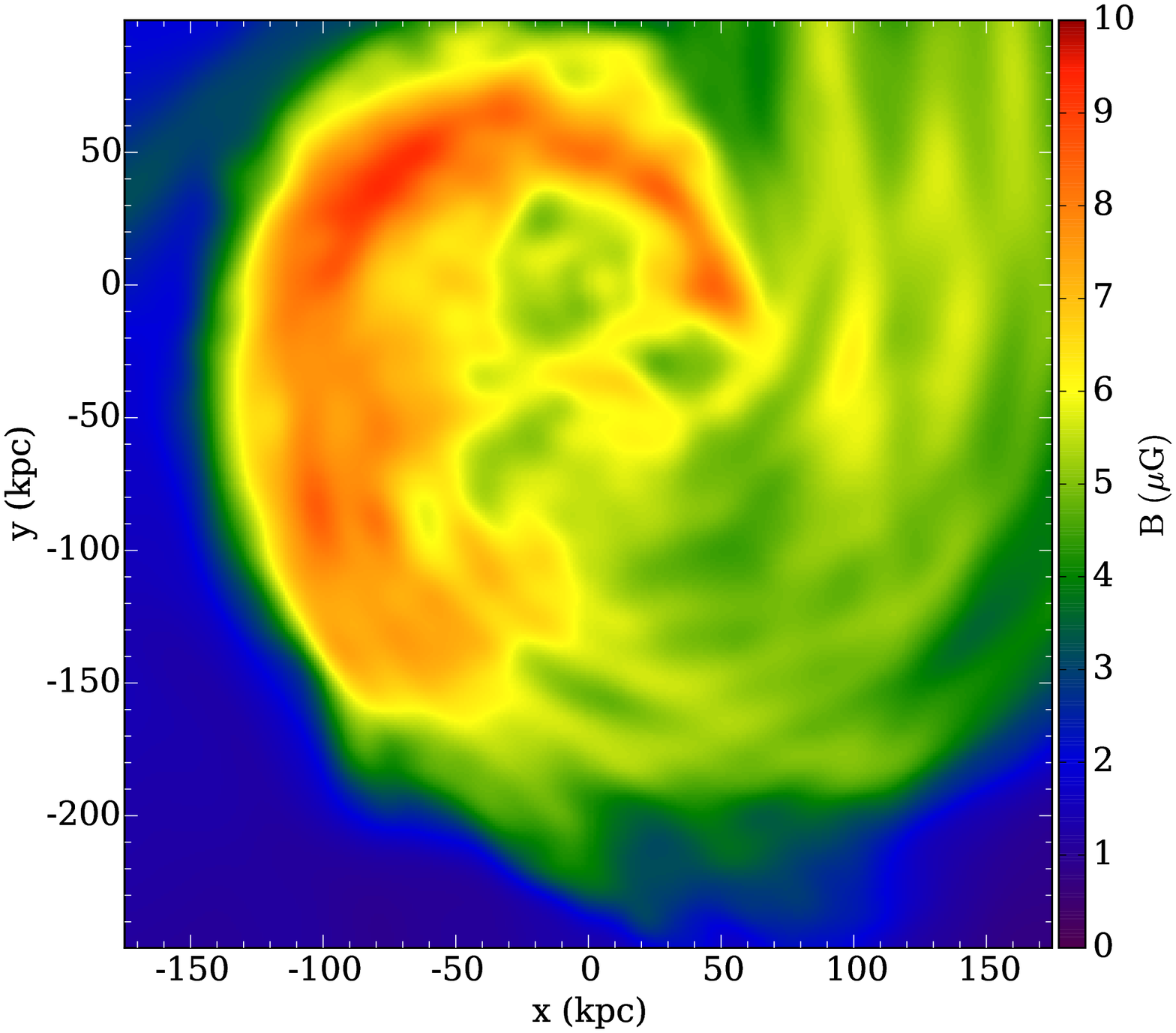}
\caption{Projections of the magnetic field strength (weighted by the density squared) of the simulation from this work, at two different resolutions. Left: $\Delta{x} = 2$~kpc. Right: $\Delta{x} = 1$~kpc. Each panel is 350~kpc on a side.\label{fig:AM06_res_test_proj}}
\end{center}
\end{figure*}

Figure \ref{fig:AM06_res_test_proj} shows projections of the magnetic field strength from the two simulations (weighted by the density squared, see Equation \ref{eqn:ss_j}). To approximate the resolution one would expect from a typical observation of a minihalo, we have smoothed each image with a Gaussian with FWHM = 10~kpc. The two maps of the magnetic field strength are very similar, with subtle differences, but the field is not significantly stronger in the higher-resolution simulation. 

\begin{figure*}
\begin{center}
\includegraphics[width=0.95\linewidth]{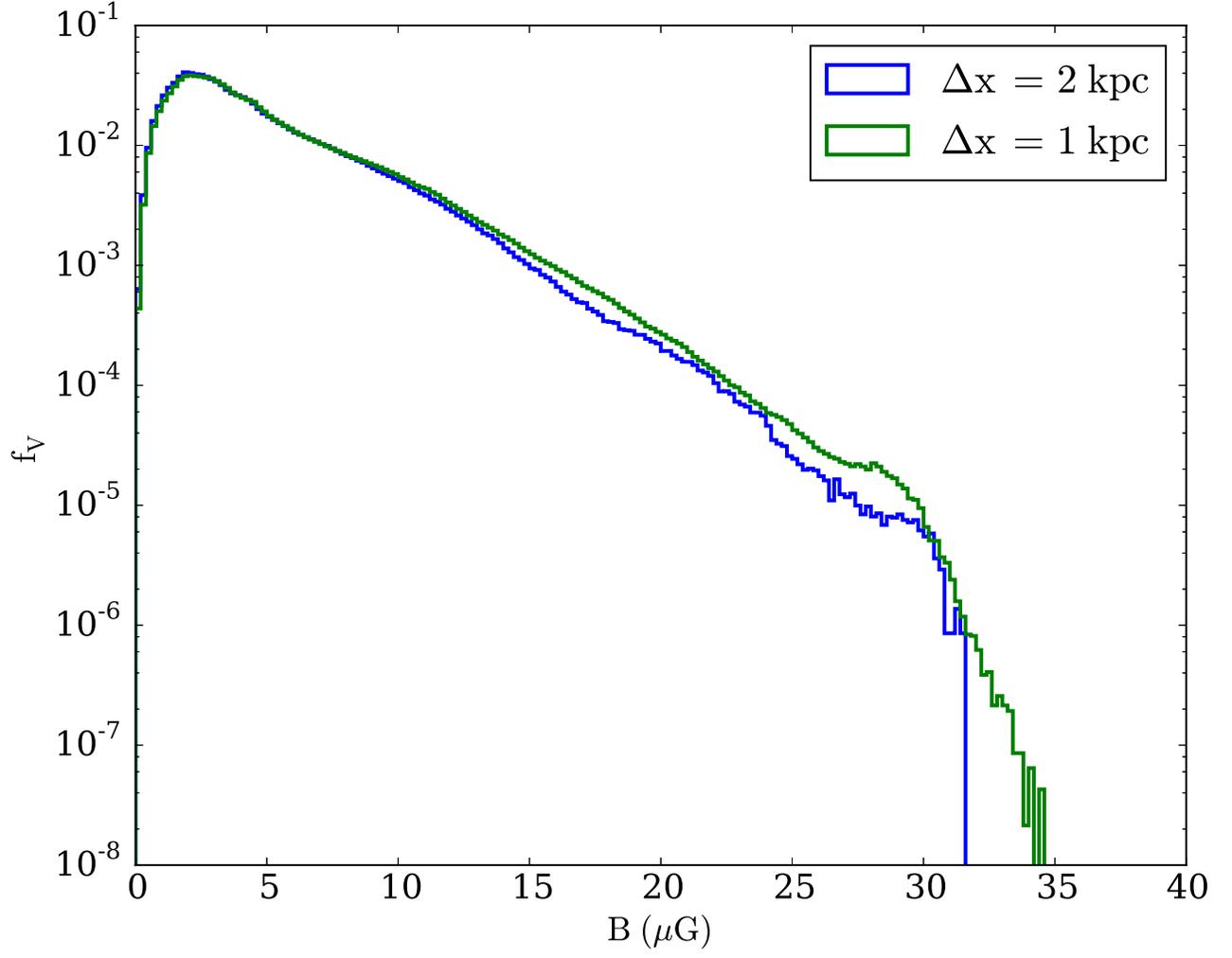}
\caption{Fraction of volume with a given magnetic field strength, for the simulation in this work with two different finest resolutions.\label{fig:AM06_histogram}}
\end{center}
\end{figure*}

The similarities in field amplification between the two simulations can also be seen in Figure \ref{fig:AM06_histogram}, which shows a binned histogram of the fraction $f_V$ of the total volume $V = (350~{\rm kpc})^3$ (centered on the cluster potential minimum) with a given magnetic field strength. The curves for the two simulations are essentially identical up to $B \sim 32~\mu{G}$, and the increase in field strength for the higher-resolution simulation does not extend much farther than this, $B \sim 35~\mu{G}$. The differences between the two simulations also occur within a very small volume, a fraction $f_V \sim 10^{-5}$ of the inner $(350~{\rm kpc})^3$ volume.

\begin{figure*}
\begin{center}
\includegraphics[width=0.95\linewidth]{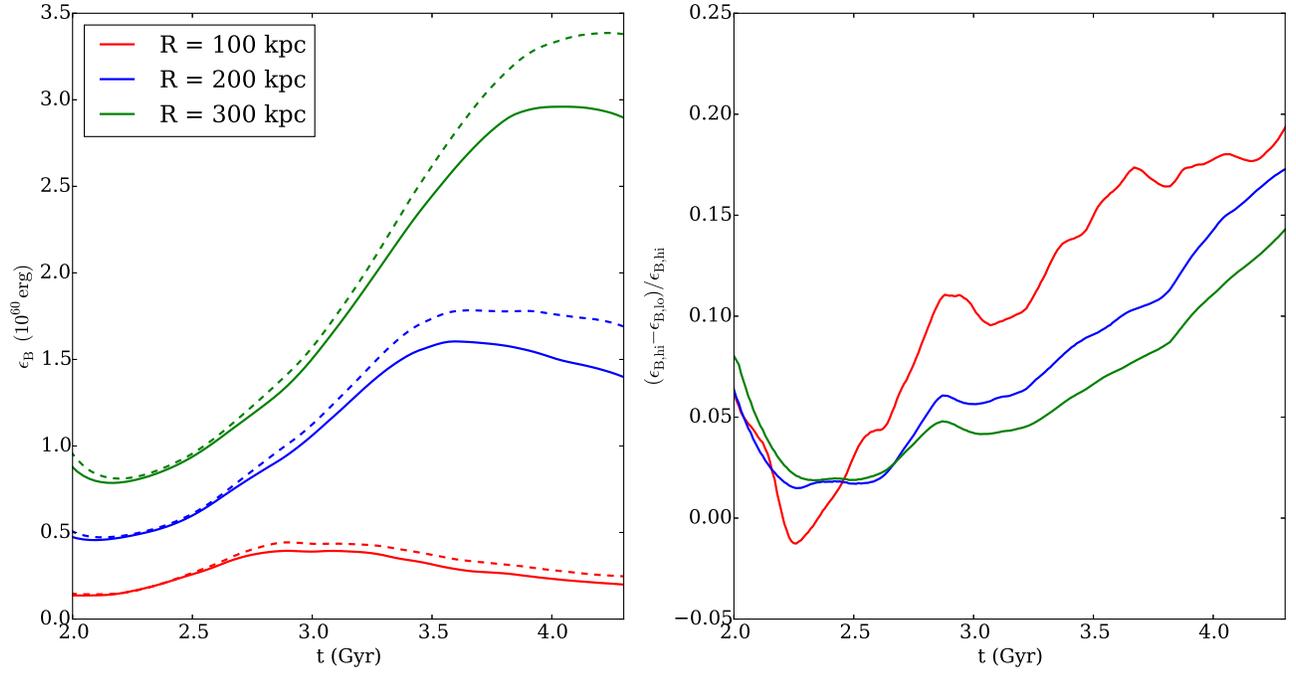}
\caption{Time evolution of the total magnetic energy $\epsilon_{\rm B}$ within three representative spherical volumes centered on the cluster potential minimum. Left: Change of the total magnetic energy within three different spherical volumes over time for the two simulations. Solid lines indicate our default simulation with $\Delta{x} = 1$~kpc, dashed lines indicate the lower-resolution version with $\Delta{x} = 2$~kpc. Right: Relative difference $(\epsilon_{\rm B,hi}-\epsilon_{\rm B,lo})/\epsilon_{\rm B,hi}$ in the total magnetic energy within the same volumes versus time.\label{fig:dBdt}}
\end{center}
\end{figure*}

Finally, we also examine the time dependence of the magnetic field amplification within the cluster core for our two different resolution cases. The left panel of Figure \ref{fig:dBdt} shows the evolution of the magnetic energy over time within three spherical volumes centered on the cluster potential minimum, defined by the radii $R =$~100, 200, and 300~kpc. In each of these volumes, we see an increase in the magnetic energy due to field amplification by the sloshing motions, followed by a decrease at later times due to the cold fronts expanding beyond the radius of each sphere (where the bulk of the increase in the magnetic field is occurring). Both simulations exhibit essentially the same time dependence with respect to the evolution of the magnetic field energy within these volumes. The right panel of Figure \ref{fig:dBdt} shows the relative difference in the magnetic energy versus time between the two simulations. In general, the higher-resolution case exhibits a greater increase in the magnetic energy over the lower-resolution case, but the difference is very small, up to $\sim$15-20\% more energy at later times. This again indicates that we have adequate resolution in our default simulation to resolve the the amplification of the magnetic field by sloshing. 

From this investigation, we conclude that for the purposes of this work the spatial resolution of our MHD simulation is adequate, since the magnetic field strength that is achieved throughout the volume of the cold fronts and on resolvable spatial scales is essentially the same between the two different simulations.

\end{document}